\documentclass{jpsj3}

\usepackage{graphicx,amsmath,amssymb,bm}

\allowdisplaybreaks 

\usepackage{color}
\definecolor{Green}{rgb}{0,0.7,0}

\newcommand{\ep}{ {\epsilon}}
\newcommand{\qTF}{ q_{\rm TF}}
\newcommand{\ET}{ $\alpha$-(BEDT-TTF)$_2$I$_3\; $}
\newcommand{\ETm}{ $\alpha$-(BEDT-TTF)$_2$I$_3$}
\newcommand{\bk}{ \bm{k}}
\newcommand{\bq}{ \bm{q}}
\newcommand{\bkD}{ \bm{k}_{\rm D}}

\newcommand{\cA}{ \chi_{\rm A}}
\newcommand{\cB}{ \chi_{\rm B}}
\newcommand{\cC}{ \chi_{\rm C}}

\newcommand{\cABC}{\chi_{\rm A+A'+B+C}}
\newcommand{\chia}{ \chi_{\alpha}}
\newcommand{\gSV}{ g(\chi^{S}+\chi^{V})}
\newcommand{\cS}{ \chi^{S}}
\newcommand{\cV}{ \chi^{V}}
\newcommand{\dmu}{ \delta \mu}
\newcommand{\tk}{ \tilde{k}}

\begin{document}

\recdate{ August 31, 2017; accepted November 15, 2017; 
published online December 28, 2017}

\title{
Effect of Long-Range Coulomb Interaction on NMR Shift in 
Massless Dirac Electrons of 
 Organic Conductor 
}
\author{
Yoshikazu Suzumura\thanks{E-mail: suzumura@s.phys.nagoya-u.ac.jp}
}
\inst{
Department of Physics, Nagoya University,   Nagoya 464-8602, Japan 
 \\
}

\hspace{8 cm}
\abst{
 The nuclear magnetic resonance (NMR) shift, $\chi_{\alpha}$, 
 at low temperatures  is examined 
  for a  massless Dirac electrons in 
  the organic conductor, $\alpha$-(BEDT-TTF)$_2$I$_3$, 
  where  $\alpha$ [= A (= A'), B, and C] denotes the sites  
 of the four molecules in the  unit cell. 
  The Dirac cone exists  within an energy of 0.01 eV 
  between the conduction and valence bands.  
  The magnetic response function is calculated by taking account of 
  the long-range Coulomb interaction  and  electron  doping. 
Calculating the interaction within  the first order in the perturbation,  
 the chemical potential is determined self-consistently, and 
   the self-energy and  vertex corrections are taken to satisfy   
   the Ward identity. 
The site-dependent $\chi_{\alpha}$ is calculated 
at low temperatures  of 
 $0.0002 < T < 0.002$  ($T$ is temperature in the unit of eV) 
by   correctly treating 
     the wave function of the Dirac cone.
 At lower  (higher) temperatures 
  the self-energy  (vertex) correction 
 of $\chia$  at all sites except for B is  dominant 
  and the sign is negative  (positive),
 while 
 the sign of the correction at the B site is always negative. 
 For moderate doping,   the shift as a function of $T$
  takes a minimum at which  $\chi_C \simeq \chi_A=\chi_{A'} > \chi_B$.  
 The relevance of the shift  to the  experiment is discussed.  
}

\maketitle

\section{Introduction} 
After the extensive studies on the electronic properties of 
low-dimensional molecular solids,\cite{Seo2004}
 a massless Dirac electron  was  found  
 in  a two-dimensional organic conductor, $\alpha$-(BEDT-TTF)$_2$I$_3$,\cite{Katayama2006_JPSJ75} 
 consisting of
  the molecule BEDT-TTF [bis(ethylenedithio)tetrathiafulvalene],
 which forms a crystal with 
 four molecules, A, A' B, and C (A = A'), in the unit cell. 
Using a tight-binding model 
  with the transfer energy estimated by 
 the extended H\"uckel method,\cite{Mori1984,Kondo2005} 
 the massless Dirac electron is described by  two valleys in the Brillouin zone 
 where  a Dirac point and Dirac cone are located 
 between the conduction and valence bands, 
  and a zero-gap state is realized owing to a three-quarter filled band.\cite{Katayama2006_JPSJ75}
The existence of the Dirac cone was verified  by 
 first-principles calculation.\cite{Kino2006} 
The effect of the Dirac cone, which causes 
 the density of states (DOS) 
 to reduce  linearly to zero  at the energy of the Dirac point,
\cite{Kobayashi2004} 
 appears  in  both electric  and magnetic properties 
  but in a different way.\cite{Kajita_JPSJ2014} 
The linear dependence of the DOS  reasonably explains 
 the conductivity  being almost constant at low temperatures,
  in addition to the conductivity at  absolute zero temperature being 
close to  the universal conductivity.
\cite{Tajima2007,Ando1998} 
The DOS of the massless Dirac cone gives  the spin susceptibility, which 
decreases linearly with decreasing temperature and shows the smallest  
  (largest) value  at site B (site C).\cite{Katayama_EPJ} 
However, the calculation in terms of the tight-binding model 
  is not enough to understand the shift of  nuclear magnetic resonance (NMR),
\cite{Takahashi2010,Hirata2012} since the deviation  of the shift 
from the linear temperature dependence is large,  
 suggesting a role of the electron correlation in the magnetic property.  
The detailed measurement of the NMR shift
\cite{Hirata2016}  
   suggested  a noticeable effect of the interaction,  
 although the relative magnitude of 
the susceptibility is compatible 
 with that of the tight-binding model.\cite{Katayama_EPJ} 
 The subsequent theoretical work studied the role of the long-range Coulomb interaction in the shift 
 on the basis of the renormalization of the velocity,
\cite{Kotov2012,Isobe2012}
  which takes account of only  
   the self-energy of the Green function. 
 Moreover, the wave function of the Dirac cone 
 must be  treated correctly, 
 since the Dirac electron in \ET is  obtained 
   by the four molecules per unit cell.
Further,   it is important to calculate the response function  
 by treating  both the self-energy and vertex corrections 
 to satisfy  the Ward identity.\cite{Ward}  
 In fact,  the vertex correction 
 of the spin-spin response function has been calculated  for  
 the on-site repulsive interaction, 
\cite{Kobayashi2013,Matsuno2017}
 where the vertex correction becomes large at high temperatures.

It the present study, we examine the NMR shift at low temperatures 
 by taking account of  
 the long-range Coulomb interaction and possible electron doping. 
  The perturbational method is applied 
 to calculate the shift  since 
 the coupling constant of the interaction 
  is small  due to  a large dielectric constant in the organic conductor, 
   as shown in the next section. 
In Sect. 2, the formulation is given where 
 the wave function is treated correctly, and  
 both self-energy and vertex corrections are calculated  
  to satisfy the Ward identity.
In Sect. 3, the solution of the chemical potential is carefully examined.
The NMR shift is  examined 
  by choosing a moderate magnitude of the interaction and the doping, and
 the result is analyzed in terms of the self-energy and vertex corrections.
In Sect. 4, we give a summary and discussion 
 on  the relevance to experiments.

\begin{figure}
  \centering
\includegraphics[width=7cm]{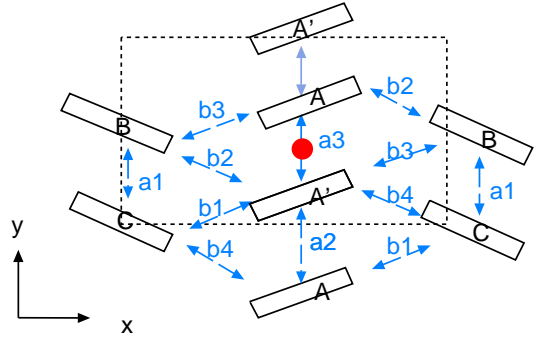}   
  \caption{(Color online)
Crystal structure  of $\alpha$-(BEDT-TTF)$_2$I$_3$ consisting of 
four molecules A, A', B and C,
 where the center of the unit cell (dotted square) 
 is taken at the middle point of A and A' (closed circle).
The transfer energies 
 between nearest-neighbor molecular sites
are given by bonds 
  $a_1, a_2, a_3, b_1, b_2, b_3$, and $b_4$.
There are also transfer energies between  next-nearest-neighbor sites 
 along the $y$-axis, and  site potentials (in the main text).
}
\label{fig:structure}
\end{figure}
 
\section{Model and Formulation}
The crystal structure of \ET is shown in 
  Fig.~\ref{fig:structure},
 which consists of four molecules ($\alpha$ = A, A', B, and C) 
  in the unit cell.
Transfer energies between nearest neighbor molecular sites are given by 
 $a1, a2, a3, b1, b2, b3$, and $b_4$.
There are also transfer energies  between  next-nearest-neighbor sites 
 along the $y$-axis, where   
  $a_{d1}$, $a_{d3}$, and $a_{d_4}$ correspond to 
 A-A, B-B, and C-C, respectively.  
 Site potentials are also added;  $p_1$, $p_2$, $p_3$, and  $p_4$ 
 act on the A, A', B,and C sites respectively, which come from the 
 mean field of the short-range repulsive interaction.  

We consider a Hamiltonian given by  
\begin{eqnarray}
H = H_0 + H_{\rm int} \; ,
\label{eq:H_model}
\end{eqnarray}
 where $H_0$ is the kinetic energy of a tight binding model 
 with  site potential $p_{\alpha}$,\cite{Kino2006,Katayama_EPJ} 
 and 
   $H_{\rm int}$ denotes the long-range Coulomb interaction 
 given by 
\begin{eqnarray}
H_0 &=& \sum_{i,j} \sum_{\alpha, \beta} \sum_{\sigma} t_{i,j;\alpha, \beta} 
    \psi_{i, \alpha,\sigma}^{\dagger} \psi_{j, \beta,\sigma} 
 + \sum_{i, \alpha} \sum_{\sigma} p_{\alpha}
     \psi_{i,\alpha,\sigma}^{\dagger} \psi_{i,\alpha,\sigma }
\; ,
\label{eq:H_0}
                                                                    \\
H_{\rm int} &=& \sum_{i, j ,\alpha, \beta} \sum_{\sigma, \sigma'}
  \frac{e^2}{|\bm{r}_{i, \alpha} - \bm{r}_{j, \beta}|}
   \psi_{i,\alpha,\sigma}^{\dagger} \psi_{j,\beta,\sigma'}^{\dagger} 
  \psi_{j,\beta,\sigma'} \psi_{i,\alpha,\sigma } \; .
\label{eq:H_int}
\end{eqnarray}
 $\psi_{i,\alpha,\sigma}^{\dagger}$ is the creation operator 
 of the electron with spin $\sigma$ for the molecular site $\alpha$ in the $i$-th unit cell, 
  forming a square lattice 
 with $N$ and $l$ being the total number of lattice sites and the lattice constant.
$t_{i,j;\alpha, \beta}$ is the transfer energy between nearest-neighbor 
  molecular sites.  
 $i$ (and $j$) denotes the  sites  of 
the unit cell forming a square lattice 
    and $\alpha$ (and $\beta$) denotes the four molecular orbitals 
 of A, A', B, and C.
Equation (\ref{eq:H_int}) denotes the long-range Coulomb interaction 
between  sites $\bm{r}_{i, \alpha}$ and $\bm{r}_{j, \beta}$.
Using the Fourier transform 
  $\psi_{\bm{k} \alpha, \sigma}
 = N^{-1/2} \sum_{j} \exp[- i \bm{k}\bm{r}_j] 
 \psi_{j,\alpha, \sigma}$, 
  where  $\bm{r}_j$ is a position vector on the square lattice, 
 Eq.~(\ref{eq:H_0}) in terms of the wave vector  $\bk = (k_x, k_y)$
is rewritten as 
\begin{equation}
H_0 = 
 \sum_{\bm{k}} \Phi_\sigma(\bm{k})^\dagger \tilde{H}_0(\bm{k}) \Phi_\sigma(\bm{k}) \; , 
\label{eq:H_0_k} 
\end{equation}
 where $\Phi_\sigma(\bm{k}) = (\psi_{\bk,A,\sigma},\psi_{\bk,A',\sigma},\psi_{\bk,B,\sigma},\psi_{\bk,C,\sigma})$
and 
$\tilde{H}_0(\bm{k})$ is  the 4$\times$4  matrix Hamiltonian given by 
\begin{eqnarray}
 \tilde{H}_0(\bm{k})
& = &
\begin{pmatrix}
  h_{\rm A}   & a   & b  & c \\
  a^* & h_{\rm A'}   & d  & e \\
  b^* & d^* & h_{\rm B}  & f \\
  c^* & e^* & f* & h_{\rm C}
\end{pmatrix} \; . 
\label{eq:H_mat}
\end{eqnarray}
 The matrix elements $a, \cdots, f$ are 
represented in terms of transfer energies  
 and the wave vector $\bm{k}=(k_x,k_y)$.
\cite{Katayama_EPJ}  
 Taking an inversion center between A and A' as the origin of the unit cell and using $\tk_x= k_x l$ and $\tk_y=k_y l$, 
these matrix elements are given by 
$h_{\rm A} = h_{\rm A'} = 2 a_{1d} \cos \tk_y + p_{\rm A }$,
$h_{\rm B}= 2 a_{3d} \cos \tk_y + p_{\rm B}$,  
$h_{\rm C} = 2 a_{4d} \cos \tk_y + p_{\rm C}$, 
$a = a_3 + a_2 {\rm e}^{i \tk_y}$,
$b = b_3 {\rm e}^{-i \tk_x/2}+ b_2 {\rm e}^{i \tk_x/2}$,  
$c = b_4 {\rm e}^{i(- \tk_x+\tk_y)/2}+ b_1 {\rm e}^{i( \tk_x+\tk_y)/2}$,  
$d = b_2 {\rm e}^{-i \tk_x/2}+ b_3 {\rm e}^{i \tk_x/2}$,  
$e =b_1{\rm e}^{i(- \tk_x-\tk_y)/2} +  b_4 {\rm e}^{i (\tk_x-\tk_y)/2}$,   
$f =a_1 ({\rm e}^{i \tk_y/2}+ {\rm e}^{-i \tk_y/2}) $. 
These transfer energies in the unit of eV are given by 
$a_1 = 0.0267$, 
$a_2 = 0.0511$, 
$a_3 = 0.0323$, 
$b_1 = 0.1241$,
$b_2 = 0.1296$, 
$b_3 = 0.0513$, 
$b_4 = 0.0512$, 
$a_{1d} = 0.0119$, 
$a_{3d} = 0.0046$, 
$a_{4d} = 0.0060$, 
$p_{\rm A} = 1.0964$,
$p_{\rm B} = 1.1475$, and 
$p_{\rm C}$ = 1.0997. 

The energy band  $\ep_\gamma(\bk)$  [$\ep_1(\bk) > \ep_2(\bk) > 
\ep_3(\bk) >\ep_4(\bk)$] 
is calculated from 
\begin{subequations}
\begin{eqnarray}
 \tilde{H}_{0}(\bk) |\gamma (\bk)> &=& \ep_{\gamma}(\bk) | \gamma (\bk)> \; ,
  \label{eq:H_eq}
 \\ 
 |\gamma (\bk)> & = & \sum_{\alpha} d_{\alpha \gamma} |\alpha> \; ,
  \label{eq:d_vector}
\end{eqnarray}
\end{subequations}
 where 
$|\gamma>$ and $|\alpha>$ denote the wave functions corresponding to 
  the energy band (eigenvalue) and the lattice site, respectively.
$ \sum_{\alpha} d_{\alpha \gamma}(\bk)^*d_{\alpha \gamma'}(\bk)
 = \delta_{\gamma, \gamma'}$ 
 and 
$ \sum_{\gamma} d_{\alpha \gamma}(\bk)^*d_{\beta \gamma}(\bk)
 = \delta_{\alpha, \beta}$. 
The component of the wave function  $d_{\alpha \gamma}(\bk)$, which is characteristic of 
 \ETm, is associated with the topological property of the wave function.
\cite{Suzumura2011} 
Although such a property also exists  in graphene,
 the novel features of the present case  arise from 
the interference effect of the four kinds of 
 $d_{\alpha \gamma}(\bk)$   
 in  the perturbational calculation   
  of   the NMR shift as shown later. 
The Dirac point, which is located 
 between the conduction  and valence bands 
 [i.e., $\ep_1(\bk)$ and $\ep_2(\bk)$ ], 
   is given by   $ \bm{k}_D/(\pi/l) = \pm (0.683,0.440)$, corresponding 
 to two valleys,  and leads to a zero gap state 
  due to the three-quarter-filled band. 

By taking account of the screening,  
Eq.~(\ref{eq:H_int}) within the random phase approximation (RPA) is rewritten as (Appendix A)
\begin{subequations}
\begin{eqnarray}                                                                 H_{\rm int}&=&  \frac{1}{Nl^2} \sum_{\bm{k}_1, \bm{k}_2, \bm{q}}
 \sum_{\alpha, \beta} \sum_{\sigma, \sigma'}
  v_{\bq,{\rm eff}} \times 
   \psi_{\bm{k}_1-\bm{q},\sigma}^{\dagger} \psi_{\bm{k}_2+\bm{q},\sigma'}^{\dagger} 
  \psi_{\bm{k}_2,\sigma'} \psi_{\bm{k}_1,\sigma } \; ,
\label{eq:H_int_k}  
 \\ 
   v_{\bq,{\rm eff}} & = & \frac{gl}{|\bq| +\qTF} 
\label{eq:veff_m}  \; ,
\end{eqnarray}
 where  
$g =  2 \pi e^2 /(l \ep$), $\ep= \ep_1 \ep_2$.
 Here the intralayer and interlayer dielectric constants are given by 
  $\ep_1=(1+ 1.43 e^2/v)$ and 
 $\ep_2 [\sim o(10)]$, respectively.
 The latter is introduced owing to the layered system 
 and is taken as a parameter   
 since  $\ep_2$ is known only for the insulating state.
\cite{Guseinov1984} 
 $e$  is the electronic charge. 
For  $l \simeq$ 10 \AA,  which is  the length of the lattice constant, 
 $2\pi e^2 /l \simeq 8.5 $ eV, 
$v/l \simeq 0.05$ eV, and  
 $e^2/v \simeq $ 27,   
 with  $v$  being the averaged velocity of the Dirac cone. 
  For $\ep_2 \simeq$  5, the coupling constant is estimated as 
 $g$ = 0.04 eV,  which is used in the numerical calculation.
Note that the dielectric constant in the present case, 
 $ \ep \simeq 200$, is much larger than   
  that of the graphene, 
   $\ep \simeq 4$,  with $e^2/v$ =  2.2.\cite{Kotov2012}  
  Since we examine the chemical potential away from the Dirac point, 
 we introduce a quantity $q_{\rm TF}(\dmu, T)$ 
  which  is the Thomas--Fermi screening constant given 
 by (Appendix A)
\begin{eqnarray}
 \qTF &=& \frac{4 e^2 / v}
{\ep (1 - \lambda^2)^{3/2}} \times \frac{|\dmu| + T}{v}
  \; ,
\label{eq:qTF_7c}
\end{eqnarray}
\end{subequations}
 where $\delta \mu = \mu -\mu_0$ and $\mu_0$ denotes $\mu$ at $g=0$ and $T=0$.    In deriving Eq.~(\ref{eq:qTF_7c}), we used a 2$\times$2 effective Hamiltonian     with  
 the tilting parameter of the Dirac cone,  $\lambda = 0.8$. 
In  Eq.~(\ref{eq:H_int_k}), we take 
 $|\bm{q}\cdot(\bm{r}_{i,\alpha}-\bm{r}_{i,\beta})| = 0$  owing to 
 the long-range Coulomb interaction. 
 We calculate $H_{\rm int}$ with a coupling constant
  $g$ (in the unit of eV) up to the first order 
 in the perturbation.

\begin{figure}
  \centering
\includegraphics[width=6cm]{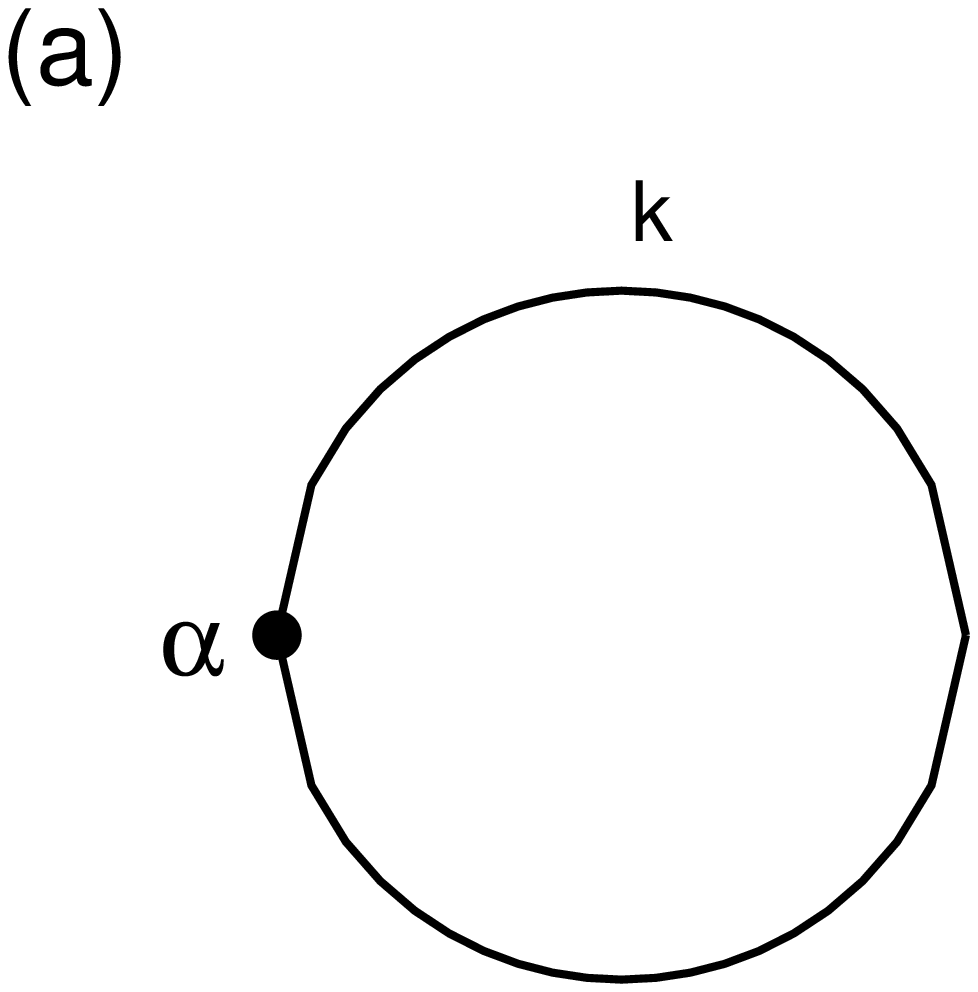}   
\includegraphics[width=6cm]{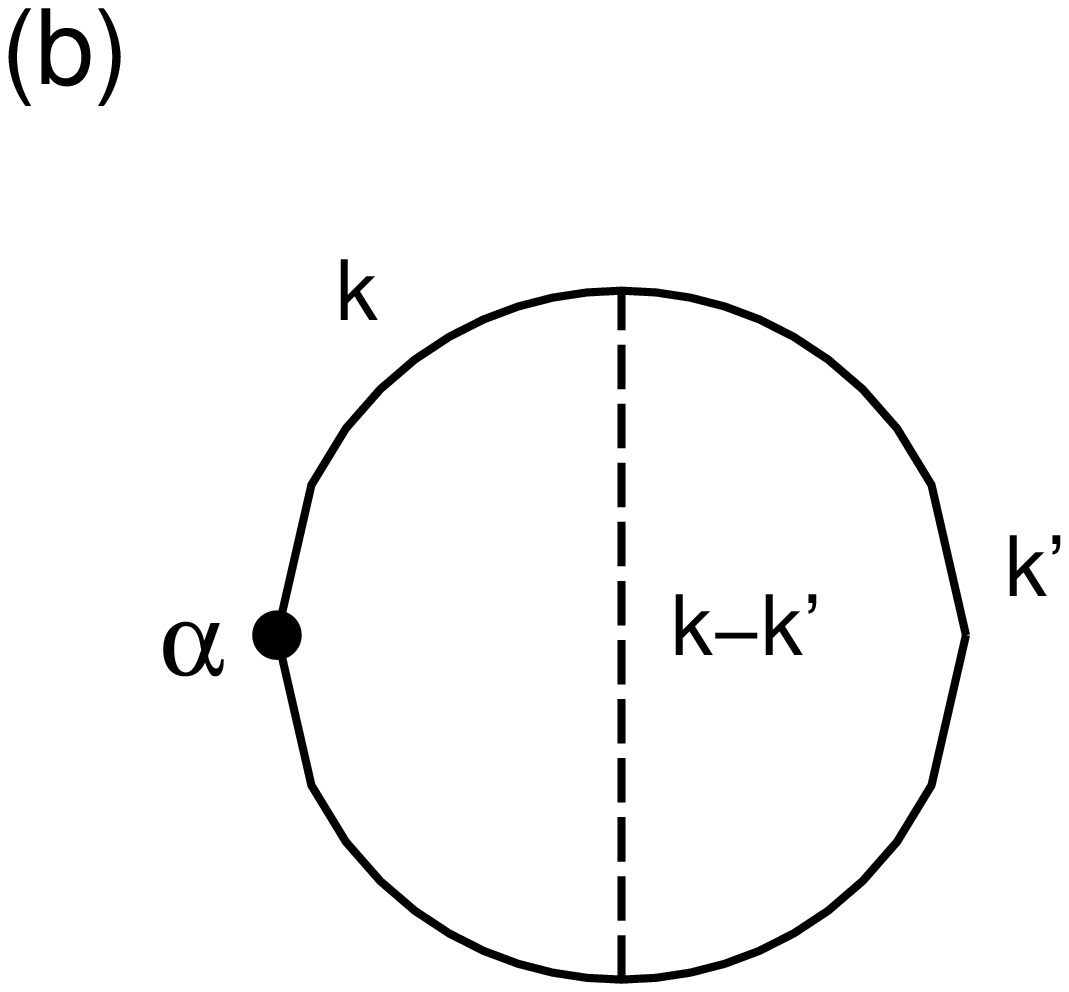}   
  \caption{
Diagram for the density of zeroth order $n^{(0)}$ (a) and  first order 
 $n^{(1)}$(b), 
where the summation of $\alpha$  is taken. 
The solid line denotes the one-particle Green function 
 ($i\omega_n + \mu  -\ep_\gamma(\bm{k}))^{-1}$, 
 where $\omega_n (= (2n+1)\pi T)$ is the Matsubara frequency 
   with $n$ being an integer. 
 The dashed line denotes 
the RPA-screened interaction $ v_{\bq,{\rm eff}}$  given by Eq.~(\ref{eq:veff_m}). 
}
\label{fig:density}
\end{figure}

 The number density per spin    
 up to the first order of the perturbation of ${H_{\rm int}}$ 
  is given by $n^{(0)} + g \; n^{(1)}$, 
 where $n^{(0)}$ and $n^{(1)}$  
 are respectively 
 shown   in  Figs.~\ref{fig:density}(a) and \ref{fig:density}(b), 
 and  are calculated as  (Appendix B) 
\begin{eqnarray}
 n^{(0)} &=& \frac{1}{N} \sum_{\bm{k}} \sum_{\gamma = 1}^{4} 
 f(\ep_{\gamma}(\bm{k}))  \; ,
\label{eq:n_0}
          \\ 
 n^{(1)} 
 &=  & - \frac{ 1}{ N^2l^2} \sum_{\bm{k}, \bm{k}'} \sum_{\gamma_1,\gamma_3} 
 \frac{1}{|\bm{k} - \bm{k}'|+\qTF} \times 
       \frac{\partial f(\ep_{\gamma_1}(\bm{k}))}
{\partial \ep_{\gamma_1}(\bm{k})} 
   f(\ep_{\gamma_3}(\bm{k}'))
       \nonumber \\
 & & 
\times \left| <\gamma_3(\bk') | \gamma_1(\bk)> \right|^2
 \; .
\label{eq:n_1}
\end{eqnarray} 
 $\sum_{\alpha} |\alpha><\alpha| 
 = \sum_{\gamma}|\gamma><\gamma|$, $f(\ep(\bk)) = 1/(\exp[(\ep(\bk)-\mu)/T] + 1)$,  and $\mu$ denotes the  chemical potential. 
 $T$ is temperature and $k_{\rm B} =1$. 
 The quantity $n^{(1)}$  is calculated as a function of $\dmu$ and $T$.

\begin{figure}
  \centering
\includegraphics[width=6cm]{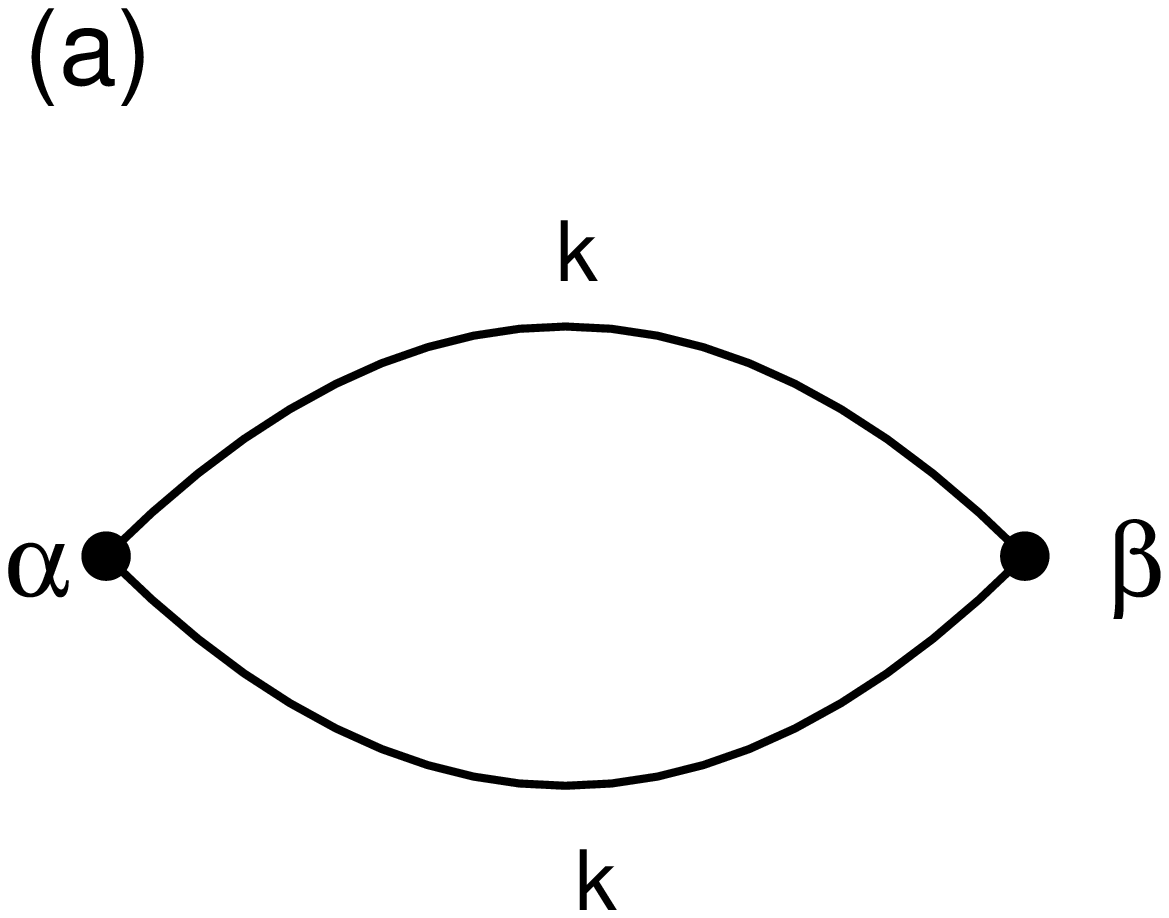}   
\includegraphics[width=6cm]{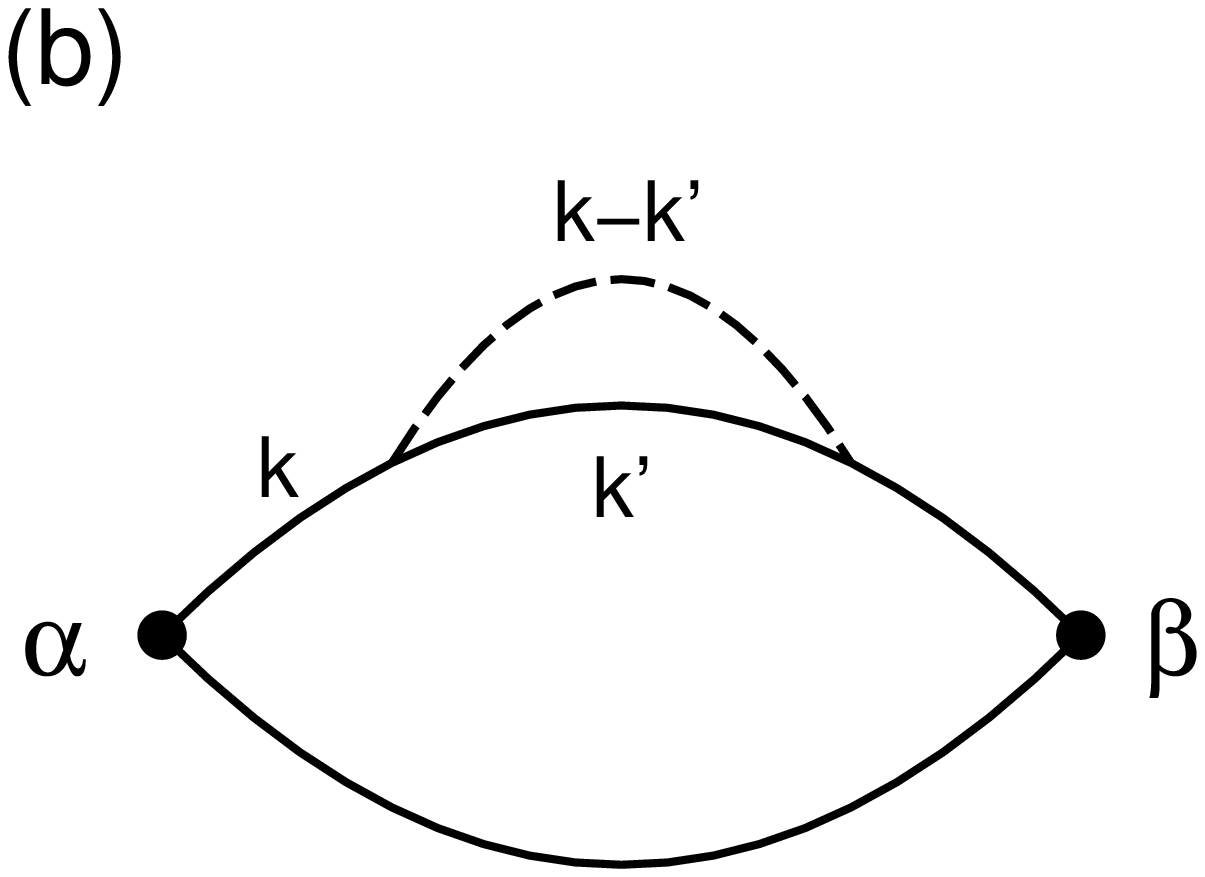}  \\
\includegraphics[width=6cm]{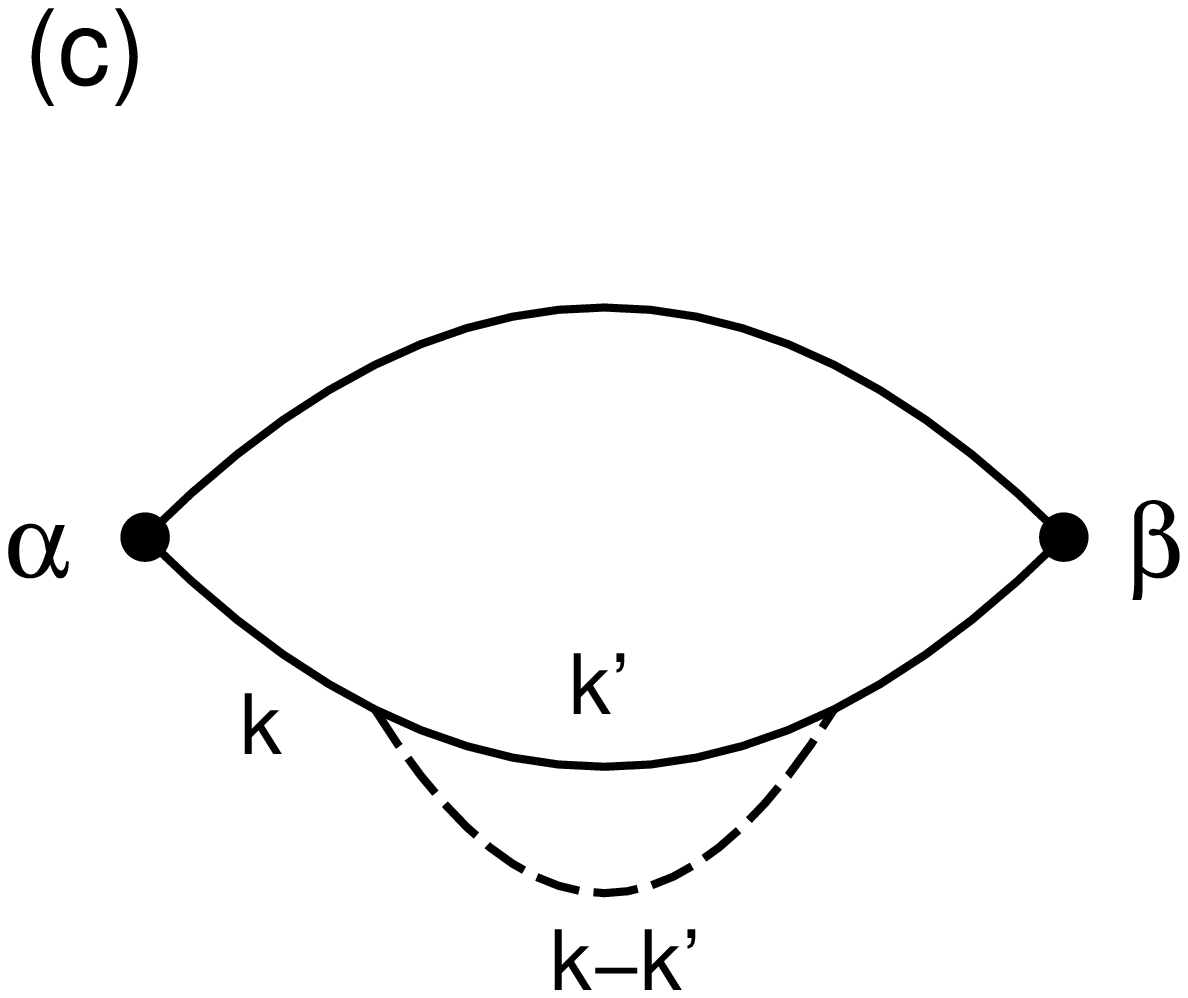}   
\includegraphics[width=6cm]{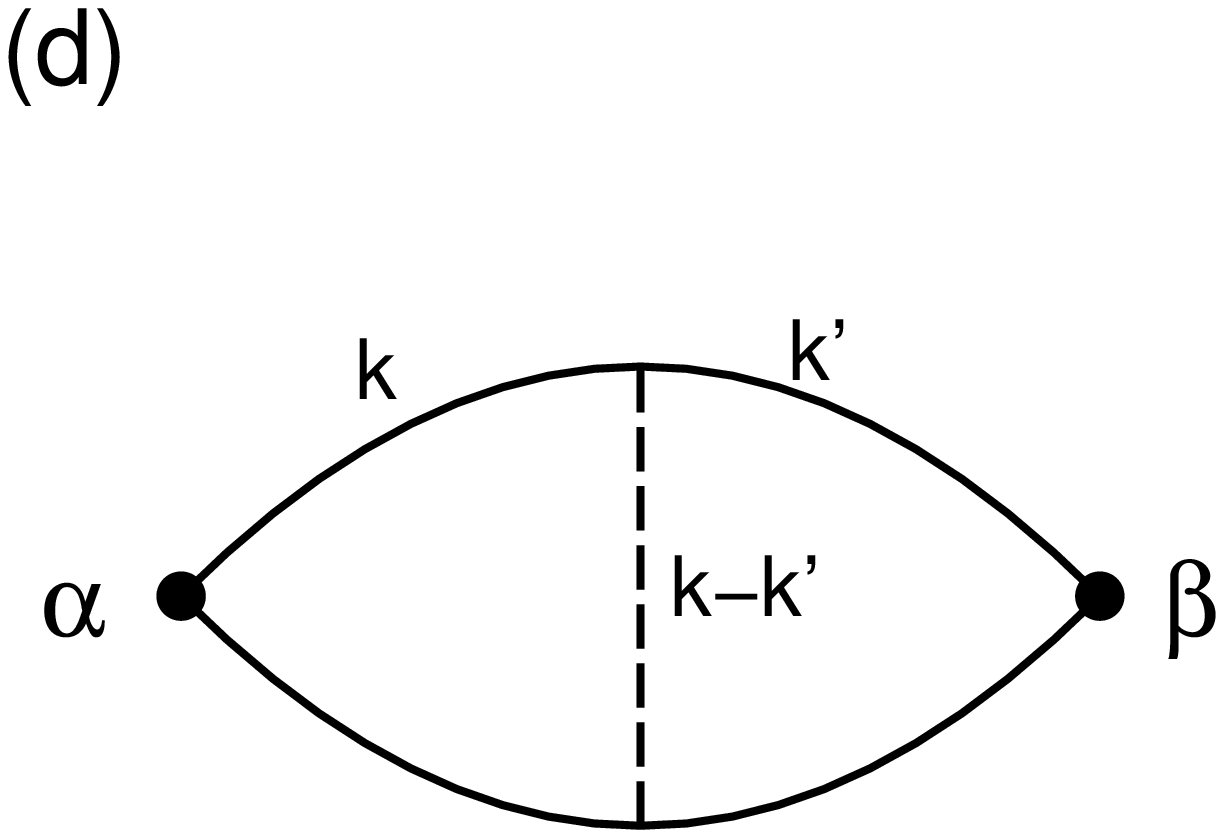} 
  \caption{
Diagram of the response function 
 for the zeroth order (a),
 the first order of the self-energy correction (b), (c), 
 and the vertex correction (d). 
Notations are the same as in Fig.~\ref{fig:density}. 
}
\label{fig:chi}
\end{figure}

 Since the number of electrons per spin and unit cell is 3,  
 the chemical potential $\mu$ is determined by  
\begin{eqnarray}
 3 + n_d &=& n^{(0)} + g \; n^{(1)}  \; ,
\label{eq:chem}
\end{eqnarray} 
 where $n_d$ denotes the doping concentration. 
For $g$ = 0, $n_d = 0$,  and  $T$=0,  $\mu$ is estimated as 
 $\mu_0$ = 1.2688, which corresponds to $\ep(\bm{k}_D)$, i.e.,   
 the energy at the Dirac point.

 We consider  an external magnetic field,  $H_{\rm ext}$,   
 applied in a direction parallel to the two-dimensional plane 
  to avoid the orbital effect of the magnetic field,  
Noting that the Zeeman energy is given by 
$ - \sum_{j} \sum_{\beta} \hat{m}_{j\beta} H_{\rm ext}$,  
the NMR shift 
 ($2 \mu_{\rm B}^2 =1$ with  $\mu_{\rm B}$ being the Bohr magneton) 
per unit cell and    at the $\alpha$ site is calculated as 
\begin{eqnarray}
 \chi_{\alpha} &= & 
 \lim_{H_{\rm ext} \rightarrow 0} \sum_{i}
 \frac{\left< \hat{m}_{i\alpha} \right> }{
NH_{\rm ext}} =   \frac{1}{2N^2}  \int_{0}^{1/T}
  \left< T_\tau \large( 
 \sum_{i}\hat{m}_{i \alpha}(0) \sum_{j \beta}\hat{m}_{j\beta}(\tau) 
\large) \right>_H    {\rm d} \tau   \; , 
\label{eq:response}
\end{eqnarray}
where  $<\cdots>_H$ denotes the average on $H$     
 in Eq.~(\ref{eq:H_model}). 
$T_\tau$ is the ordering operator of the imaginary time $\tau$,
$\hat{m}_{j\alpha} = 
 \hat{n}_{j\alpha \uparrow} -\hat{n}_{j\alpha \downarrow},$ and 
 $\hat{n}_{j\alpha \sigma} 
 =  \psi_{j \alpha \sigma}^{\dagger} \psi_{j \alpha \sigma}$.
It is crucial that  the  shift at the $\alpha$ site 
 is affected  not only by the same kind of  molecule but also 
 by the different kinds of  molecules 
 due to  four molecules per unit cell.
The shift up to first order in terms of the perturbation  is given by  
\begin{eqnarray}
 \chi_\alpha & \simeq &  \chi_\alpha^{(0)} + g \; \chi_\alpha^{S} + g \;\chi_\alpha^{V} \; ,
\label{eq:chi_total}
\end{eqnarray}
 which is calculated using a response function 
in terms of the Green function.\cite{Abrikosov}
The first term denotes the zeroth order
  given by Fig.~\ref{fig:chi}(a).
 The second term of Eq.~(\ref{eq:chi_total}) 
   is the self-energy correction of the first order 
     given by Figs.~\ref{fig:chi}(b) and \ref{fig:chi}(c).
 The third term of Eq.~(\ref{eq:chi_total}) 
   is the vertex correction of the first order 
     given by Fig.~\ref{fig:chi}(d). 
It should be noted that,  in addition to the second and  third terms,  
 another contribution called  the A--L term\cite{Aslamazov} 
is generally required  to satisfy the Ward identity,\cite{Ward} 
 as shown for the fluctuation conductivity.  
 However  Figs.  \ref{fig:chi}(b),  \ref{fig:chi}(c), and \ref{fig:chi}(d) 
 are enough in the present case of the magnetic field 
 due to the cancellation by  the summation of 
 $\beta$ in  $\sum_{j \beta}\hat{m}_{j\beta}(\tau)$
 of  Eq.~(\ref{eq:chi_total}).

 The response function of the zeroth order  
  is calculated as (Appendix C)
\begin{eqnarray}
\chi_{\alpha}^{(0)} &=& \sum_{\beta} \chi_{\alpha \beta}^{0}
  = - \frac{1}{N}\sum_{\bm{k},\gamma} \frac{\partial f(\ep_{\gamma}(\bm{k}))}
 {\partial \ep_{\gamma}(\bm{k})}
   d_{\alpha \gamma}^{*}(\bm{k})d_{\alpha \gamma}(\bm{k}) \; , 
  \label{eq:chi_0}
\end{eqnarray}                       
which is rewritten as 
\begin{eqnarray}
\chi_{\alpha}^{(0)} &=& 
    - \int_{-\infty}^{\infty} {\rm d} \omega 
         \frac{\partial f(\omega)}{ \partial \omega} 
   D_{\alpha}(\omega) \; ,  
  \label{eq:chi_01}
  \\ 
D_\alpha(\omega) &=& \frac{1}{N} \sum_{\bk} \sum_{\gamma}
 \delta (\omega - \ep_{\gamma}(\bk)) 
 d_{\alpha \gamma}^{*}(\bm{k})d_{\alpha \gamma}(\bm{k})
\; .
  \label{eq:DOS}
\end{eqnarray}
 $D_\alpha(\omega)$ denotes the local DOS per spin and unit cell, 
 the total DOS is   $D({\omega}) = \sum_{\alpha} D_\alpha(\omega)$,  and 
  $\int {\rm d} \omega D(\omega) = 4$.
At low temperatures, for which the numerical calculation is performed 
in the next section, we obtain $\chi_{\alpha}^{(0)} \propto T$ due to 
  $D_{\alpha}(\omega) \propto |\omega|$.

 Performing a summation over $\beta$ in Figs.~\ref{fig:chi}(b) and 
 \ref{fig:chi}(c),  
 the second term of Eq.~(\ref{eq:chi_total}) is calculated as (Appendix C) 
\begin{eqnarray}
g \chi_\alpha^{S} & =&  \frac{g}{2 N^2l^2}
   \sum_{\bm{k},\bm{k}'} \sum_{\gamma_1, \gamma_2,\gamma_3} 
\frac{1}{|\bm{k}-\bk'|+\qTF} \times
  \frac{1}{\ep_2 -\ep_1}  \left( 
   \frac{\partial f_2}{ \partial \ep_2}
  - \frac{\partial f_1}{ \partial \ep_1}
  \right)   
 \times  f(\ep_{\gamma_3}(\bm{k}'))
         \nonumber \\
 &  \times &
\left( <\gamma_1(\bk) | \alpha> 
 <\alpha | \gamma_2(\bk)> 
 <\gamma_3(\bk') | \gamma_1(\bk)> 
 <\gamma_2(\bk) | \gamma_3(\bk')> + (c.c.) \right)
         \; ,
  \label{eq:chi_self}
\end{eqnarray}
 where $f_1=f(\ep_1)$, $f_2=f(\ep_2)$, $f_4=f(\ep_4)$, 
 $\ep_1= \ep_{\gamma_1}(\bm{k})$, $\ep_2=\ep_{\gamma_2}(\bm{k})$, and 
 $\ep_4= \ep_{\gamma_4}(\bm{k}-\bm{q})$. 
Performing a summation over $\beta$ in Fig.~\ref{fig:chi}(d),  
 the third  term of Eq.~(\ref{eq:chi_total}) is calculated as (Appendix C)
\begin{eqnarray}
g  \chi_\alpha^{V} & =& 
   \frac{g}{N^2l^2} \sum_{\bm{k},\bm{k}'} \sum_{\gamma_1, \gamma_2,\gamma_3} 
 \frac{1}{|\bm{k} - \bk'| + \qTF} \times 
  \frac{f_1-f_2}{\ep_1 -\ep_2} \times  \frac{\partial f_3}{\partial \ep_3} 
         \nonumber \\
 & & \times 
 <\gamma_1(\bk) | \alpha> 
 <\alpha | \gamma_2(\bk)> 
 <\gamma_3(\bk') | \gamma_1(\bk)> 
 <\gamma_2(\bk) | \gamma_3(\bk')> 
         \; ,
 \label{eq:chi_vertex}
\end{eqnarray}
 where $f_1=f(\ep_1)$, $f_2=f(\ep_2)$, $f_3=f(\ep_3)$, 
 $\ep_1= \ep_{\gamma_1}(\bm{k})$, $\ep_2=\ep_{\gamma_2}(\bm{k})$, and 
 $\ep_3= \ep_{\gamma_3}(\bm{k}')$. 


\section{ NMR Shift }

\subsection{Chemical potential}
The  chemical potential $\dmu$ 
  is calculated  self-consistently 
     using  Eq.~(\ref{eq:chem}), which is rewritten as 
\begin{eqnarray}
   n_{\rm hole} + n_d = g \; n^{(1)} \; ,
\label{eq:SCE}
 \end{eqnarray}
 where $ n_{\rm hole} =3 - n^{(0)}$. 
Equation (\ref{eq:SCE}) gives $\dmu$ as a function of $T$, $n_d$, 
and  $g$, i.e., $\dmu(T, n_d, g) $.
In order to obtain $\dmu$ as a function of $T$, $n_d$, and  $g$, 
Eqs.~(\ref{eq:n_0}) and (\ref{eq:n_1}) (i.e., $n^{(0)}$ and  $n^{(1)}$)
 are calculated  as a function of  $\dmu$ and $T$,  
   where  $\delta \mu = \mu -  \mu_0$ with $\mu_0$ given by  
 $\ep(\bm{k}_D)$ at $T=0$. 

First we examine $\dmu$ at $T$=0.
Using the effective 2$\times$2 Hamiltonian of the Dirac cone (Appendix A)
  with velocity $v$ and tilting parameter $\lambda$,
 Eq.~(\ref{eq:n_0}) is calculated as 
\begin{eqnarray}
   n_{\rm hole} &=& - {\rm sgn}(\dmu ) 
      \frac{\dmu^2l^2}  {4 \pi v^2}
      \frac{1}{(1-\lambda^2)^{3/2}} 
         \label{eq:n_T=0} \; .
 \end{eqnarray}
In the present case of $\lambda$ = 0.8  and $v/l \simeq$ 0.05,   
  $n_{\rm hole} =- {\rm sgn}(\dmu ) C_0 \dmu^2$ with 
 $C_0 \simeq$ 150 (eV)$^{-2}$.
 Equation (\ref{eq:n_1}) is also estimated as  
  $n^{(1)} = C_1 |\dmu|$ with $C_1 \simeq 12$ (eV)$^{-2}$ (Appendix B).  
Substituting these values into  
Eq.~(\ref{eq:SCE}), $\dmu$ is obtained as follows.
For $n_d = 0$, $\dmu = - g(C_1/C_2) (<0)$, while 
 $\dmu = (-g C_1 + \sqrt{(gC_1)^2 + 4 C_0 n_d})/(2C_0) (>0)$
 for $n_d > (gC_1)^2/(4C_0)$.
In the range of  $0 < n_d < (gC_1)^2/(4C_0)$,  
 there are three kinds of solutions, 
 $[
\dmu = 
(-g C_1 + \sqrt{(gC_1)^2 + 4 C_0 n_d})/(2C_0)$  and  
$(-g C_1 \pm \sqrt{(gC_1)^2 - 4 C_0 n_d})/(2C_0)
]$, 
where we take the smallest one, 
 $\dmu = (-g C_1 - \sqrt{(gC_1)^2 - 4 C_0 n_d})/(2C_0) (<0)$,
 in order to obtain a solution connected continuously to that of  $T$=0. 
 Thus,  a first-order transition occurs  
  at  $n_d =  (gC_1)^2/(4C_0)$, 
 where  the sign of the chemical potential $\delta \mu$ changes from 
  negative to  positive  with decreasing $g$ or increasing $n_d$.

Here we mention the state given by $\delta \mu < 0$  
for  $n_d = 0$. 
 Since $\delta \mu < 0$ gives  $n^{(1)} > 0$ from  Eqs.~(\ref{eq:n_1}) and  (\ref{eq:n_T=finite}), the chemical potential is located at the valence band with  $\ep_2 (\bk) - \mu_0 = \delta \mu (<0)$.
 This implies the emergence of an excess electron density at $\bk$ with $\ep_2(\bk)=\mu$ 
 in the valence band, which has the effect of reducing  the chemical potential to keep the total number of  filled electrons. 
 Thus, holes exist in the valence band  below the Dirac point (i.e., the  valley of the Dirac cone) even for  $n_d$ = 0. 

\begin{figure}
  \centering
\includegraphics[width=7cm]{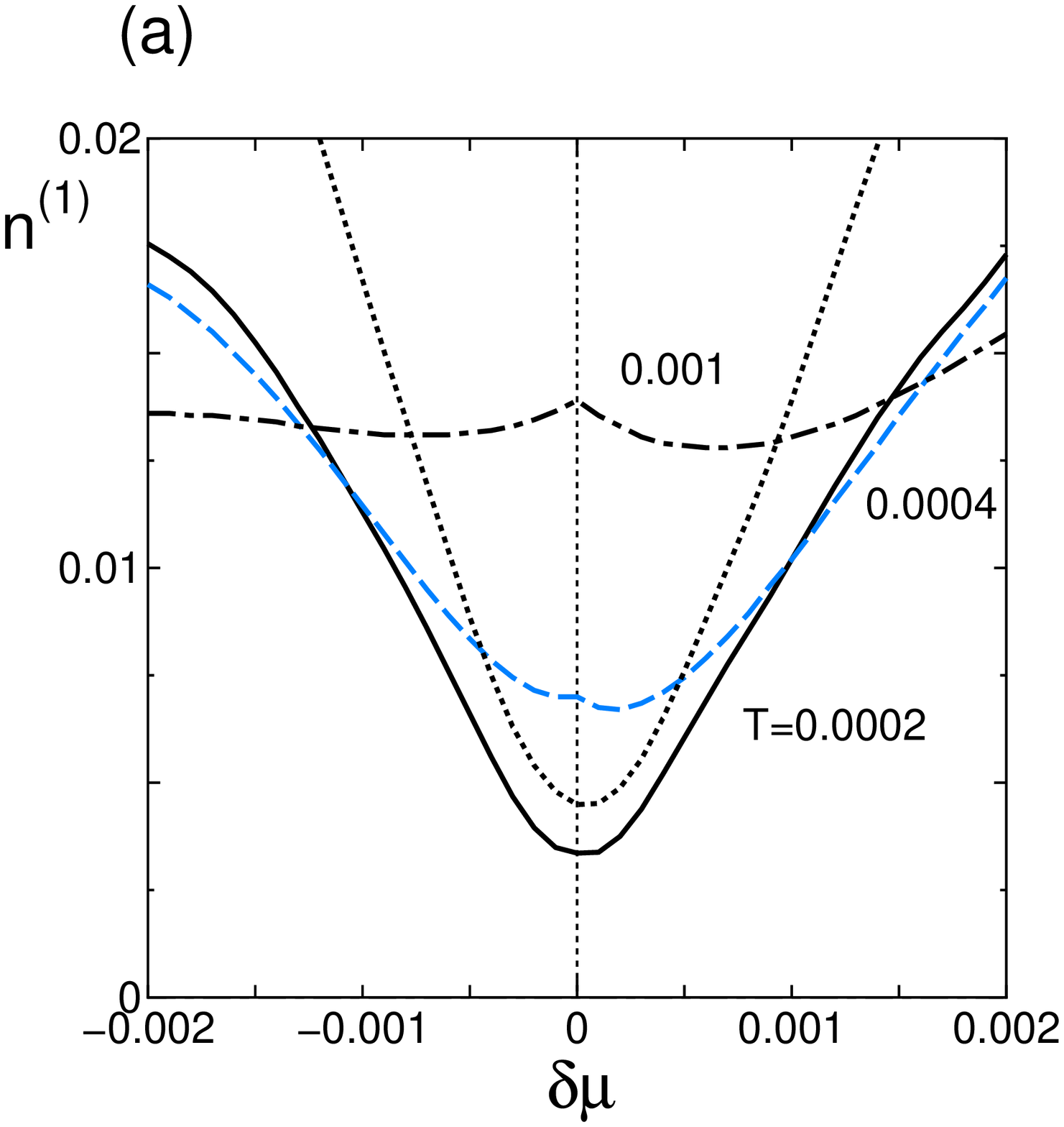}   
\includegraphics[width=7cm]{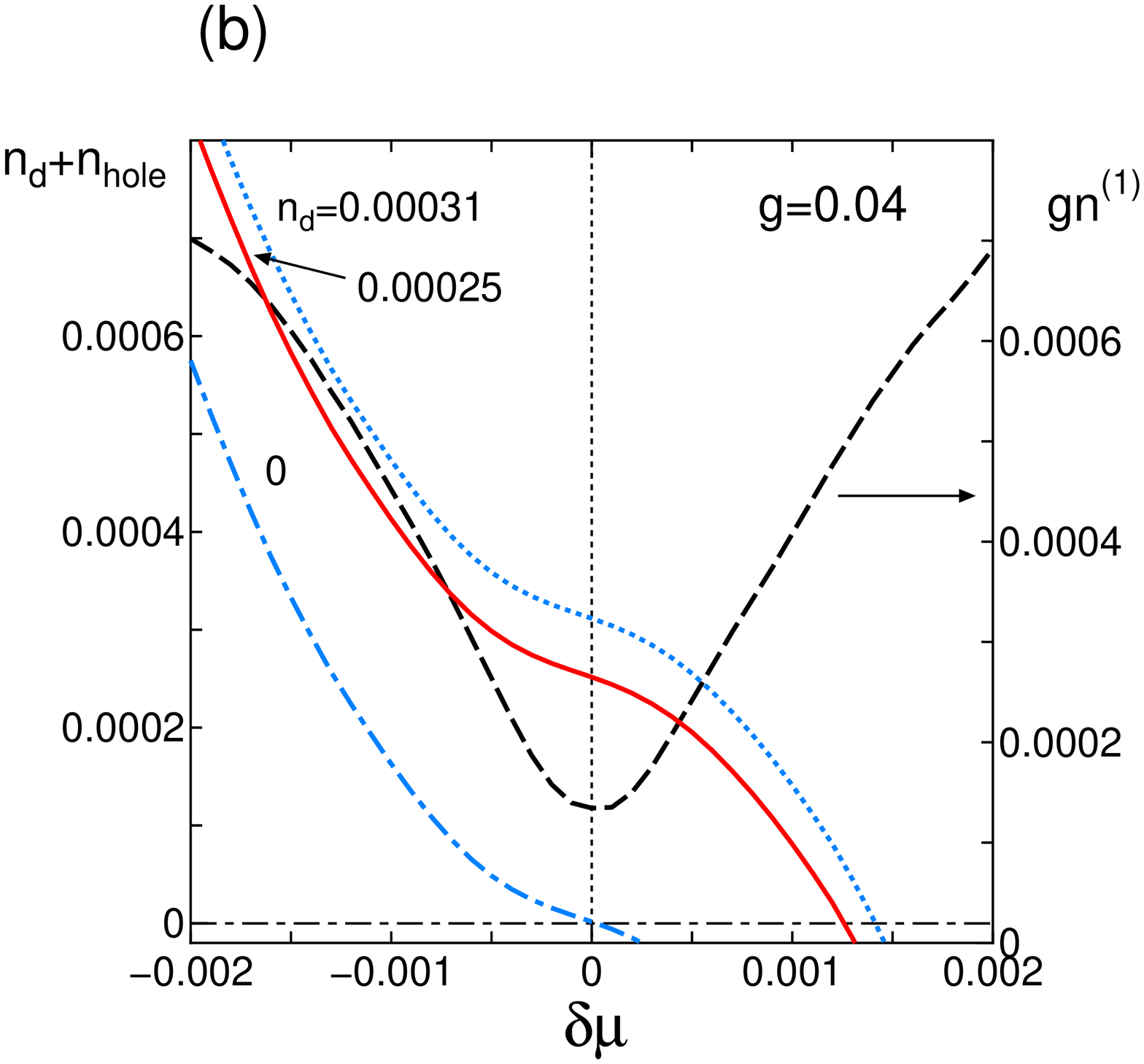}   
  \caption{(Color online)
 (a) $\delta \mu$ ($= \mu - \mu_0$) dependence of $ n^{(1)}$  
  with fixed $T$ = 0.001,  0.0004,  and 0.0002, 
 where the dotted line denotes  $ n^{(1)}$ with $q_{TF}=0$ at $T$=0.0002.
 (b) $\delta \mu$  dependence of 
 $n_{\rm hole}+n_d$ and  $g n^{(1)}$ 
 for $T$ = 0.0002 with $g$=0.04, 
 where $n_{\rm hole} = 3 - n^{(0)}$ and 
 $n_d$ denotes the doping concentration.
 $\mu_0 (=1.2688)$ denotes the chemical potential for $T=0$ and $n_d = 0$ 
 in the absence of interaction.  
The intersection  gives a solution for $\delta \mu$, where the lowest one 
 is taken when there are  many solutions. 
}
\label{fig:fig4}
\end{figure}

Next we examine $\dmu$ for $T \not= 0$, 
 which is calculated numerically from  Eq.~(\ref{eq:chem}). 
When there is more than one solution, we choose the smallest one 
 in order to be  consistent with that of $T$ = 0. 
Figure \ref{fig:fig4}(a) shows the $\dmu$ dependence  
 of  $ n^{(1)}$  for fixed  $T$ = 0.0002 (solid line), 0.0004 (dashed line) 
 and 0.001 (dot-dashed line).
The quantity $n^{(1)}$ is positive, where  $n^{(1)} = 0$ at $T$=0,  
  and 
 $n^{(1)}$ at low temperatures is proportional to $T$ 
  due to the factor 
 $-\partial f(\ep_{\gamma_1}(\bm{k}))/\partial \ep_{\gamma_1}(\bm{k})$. 
It is found that  $ n^{(1)}$ as a function of $\dmu$  shows  
  $ n^{(1)}(\dmu) - n^{(1)}(0) \propto \dmu^2$  for small $\dmu$, 
 although there is a slight deviation from the symmetric behavior 
 and a slight maximum at $\dmu = 0$. 
  In order to see  the suppression of 
  $n^{(1)}(\dmu)$ by the screening,
   $ n^{(1)}$ with $q_{TF}=0$ at $T$=0.0002 
     (dotted line)   is compared with    the solid line.   
Figure \ref{fig:fig4}(b) shows the $\dmu$ dependence  
of    $n_d+n_{\rm hole}$ and $ g n^{(1)}$ with $g$ = 0.04 for $T$=0.0002, 
 where  the intersection  gives a solution of $\dmu$. 
Thus, the chemical potential  $\dmu$  is calculated 
  self-consistently  for fixed  $T$, $n_d$, and $g$.  
The solution of $\dmu$ is a single value for $n_d$ =0.00031 and 0. 
 For $n_d$=0.00025,
  there are three solutions and the lowest $\dmu$ is chosen 
  as shown  for $T$=0.

\begin{figure}
  \centering
\includegraphics[width=7cm]{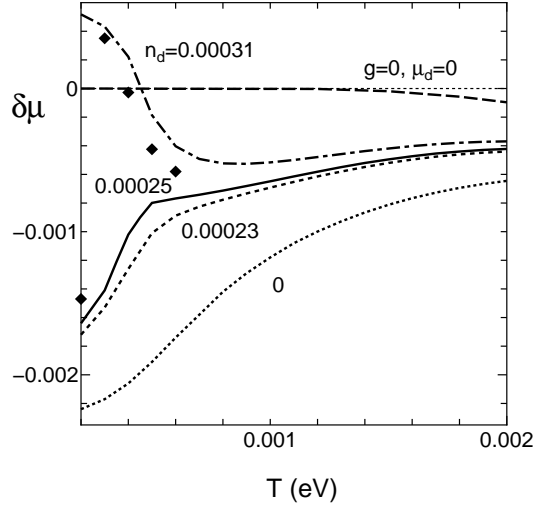}   
  \caption{
$T$ dependence of $\delta \mu$ for $g$ = 0.04 with 
 fixed $n_d$ = 0.00031, 0.00025, 0.00023, and 0,   which 
are obtained from  $n_d + n_{\rm hole} = g n^{(1)}$. 
The symbols (diamonds)  correspond to $\dmu$ for 
$n_d$ =  0.00028.
The dashed line ($g$=0) denotes $\delta \mu$ for $g=0$ and $n_d=0$.
}
\label{fig:fig5}
\end{figure}
 
Figure \ref{fig:fig5} shows the  $T$ dependence of $\dmu$ 
 with  some choices of $n_d$ for $g$=0.04,  
where there are the following three types of $T$ dependence of 
$\dmu$,  depending on $n_d$. 
 For large $n_d$ (= 0.00031),
 there is a crossover from  $\dmu >0$ 
 to $\dmu < 0$ with increasing $T(>0.0001)$. 
$\dmu$ takes a minimum above the temperature corresponding to $\dmu = 0$. 
 For small $n_d$ (=0.00023 and 0), 
$\dmu < 0$ exists for arbitrary $T$  and 
  $\dmu$ increases monotonically with increasing $T$.
In the region of $0.0026 < n_d < 0.00031$ 
(for example, $n_d$ =0.0028 (diamonds)), 
 $\dmu$ jumps from $\dmu < 0$ to $\dmu > 0$ with increasing $T$ $(> 0.0002)$, 
  while such a jump diminishes for $T > 0.005$. 
Based on more precise calculation, we find that 
 the jump of $\dmu$ occurs at 
 $(n_d, T) \simeq$ (0.00032, 0), (0.00031, 0.0002), (0.00028, 0.0003),
  (0.00026, 0.0004), 
 forming a line of the boundary between  $\dmu < 0$ and $\dmu > 0$, 
which terminates before  $ T \simeq 0.0005$.
For simplicity, the present paper does  not treat  such a region  where 
 a first-order transition occurs at low temperatures 
 ($T < 0.0005$). 
Using $\dmu$ of  Fig.~\ref{fig:fig5} with a moderate choice of  $n_d$, 
 we examine the NMR shift $\chi_{\alpha}$ in the next section 
  to obtain a similar result  to that of an experiment at low temperatures. 
The choice of $n_d$ is discussed in  Sect. 4.

\subsection{NMR shift}
 The numerical calculation of the NMR shift is performed as follows. 
The  zeroth-order term given by Eq.~(\ref{eq:chi_0})
   is calculated  by dividing the summation into $n=200$ segments for 
     the axes of $k_x$ and $k_y$ in the first Brillouin zone. 
 Equation (\ref{eq:H_int_k}) is calculated in 
   the reduced region consisting of  two valleys  
      around the Dirac point $\pm \bkD$. 
 In order to examine the effect of the interaction 
 at low temperatures of $T < 0.002$, 
    the calculation of  Eqs.~(\ref{eq:n_1}), (\ref{eq:chi_self}), and
    (\ref{eq:chi_vertex}) is performed  by choosing 
      $|\bk \pm  \bkD|/\pi < 0.1$ with 40 segments.  
This choice is reasonable 
 since  the change by  $|\bk \pm  \bkD|/\pi < 0.14$ 
 is  less than 10$\%$. 
 The NMR shift is examined in the range of $0.0002 < T < 0.002$
 due to the limited number of segments.

\begin{figure}
  \centering
\includegraphics[width=7cm]{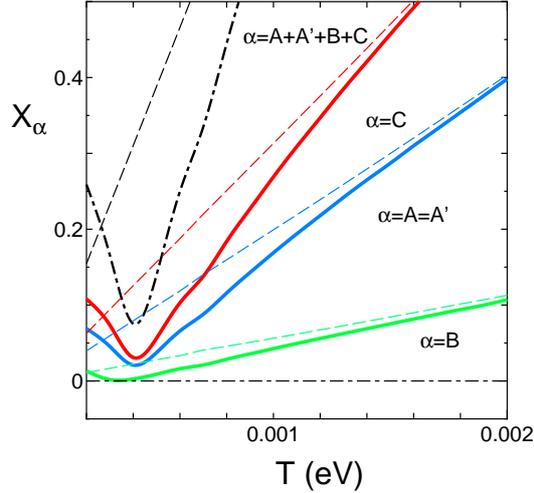}   
  \caption{(Color online)
$T$ dependence of NMR shift $\chi_{\alpha}$ 
 with $\alpha$ = A (=A'), B, and C  
for $g$ = 0.04 and $n_d=0.00025$.
 The dashed line denotes $\chi_{\alpha}^{(0)}$ 
 for $g=0$, $n_d=0.0$,  and  $\delta \mu=0$. 
Note that $\chi - \chi^{(0)}$ corresponds to 
the  sum of  the self-energy and vertex 
 corrections,  where $\chi = \cABC$. 
}
\label{fig:fig6}
\end{figure}
Using the chemical potential $\delta \mu$ obtained in Fig.~\ref{fig:fig5},
 we calculate  Eq.~(\ref{eq:chi_total}) to examine 
 the $T$ dependence of the NMR shift.
 Figure \ref{fig:fig6} shows the  $T$ dependence of $\chia$  with $g=0.04$ 
 and $n_d = 0.00025$, where 
      $\alpha$  =   A(=A'), B, and C denotes the shift 
       for the  respective site  and 
        $\alpha$ = A+A'+B+C denotes  the sum  of the shift.
It is noticed that the relation  $\cC > \cA >\cB $ still holds even
   in the presence of the interaction.
The dashed line denotes $\chi_{\alpha}^{(0)}$, i.e., the shift in the case of 
   $g=0$, which is proportional to  $T$.\cite{Katayama_EPJ} 
Compared with $\chi_{\alpha}^{(0)}$, 
  $\chia$ exhibits a noticeable reduction, i.e.,  
  suppression,  which  comes from $\gSV$ ($<0$). 
At $T \simeq$ 0.0005, $\cC$ and $\cA$ show a minimum and  
 $\cC \simeq \cA$,  
 while  $\cB$ reduces almost to zero. 
There is an enhancement of $\cC$ and $\cA$ at low temperatures 
 due to the finite  $|\dmu|$,  which increases  $\chia^{(0)}$.
The suppression of $\chia$ becomes large for larger  $g$ since 
   $g\cS$  and $g\cV$ are mainly proportional to $g$.

\begin{figure}
  \centering
\includegraphics[width=7cm]{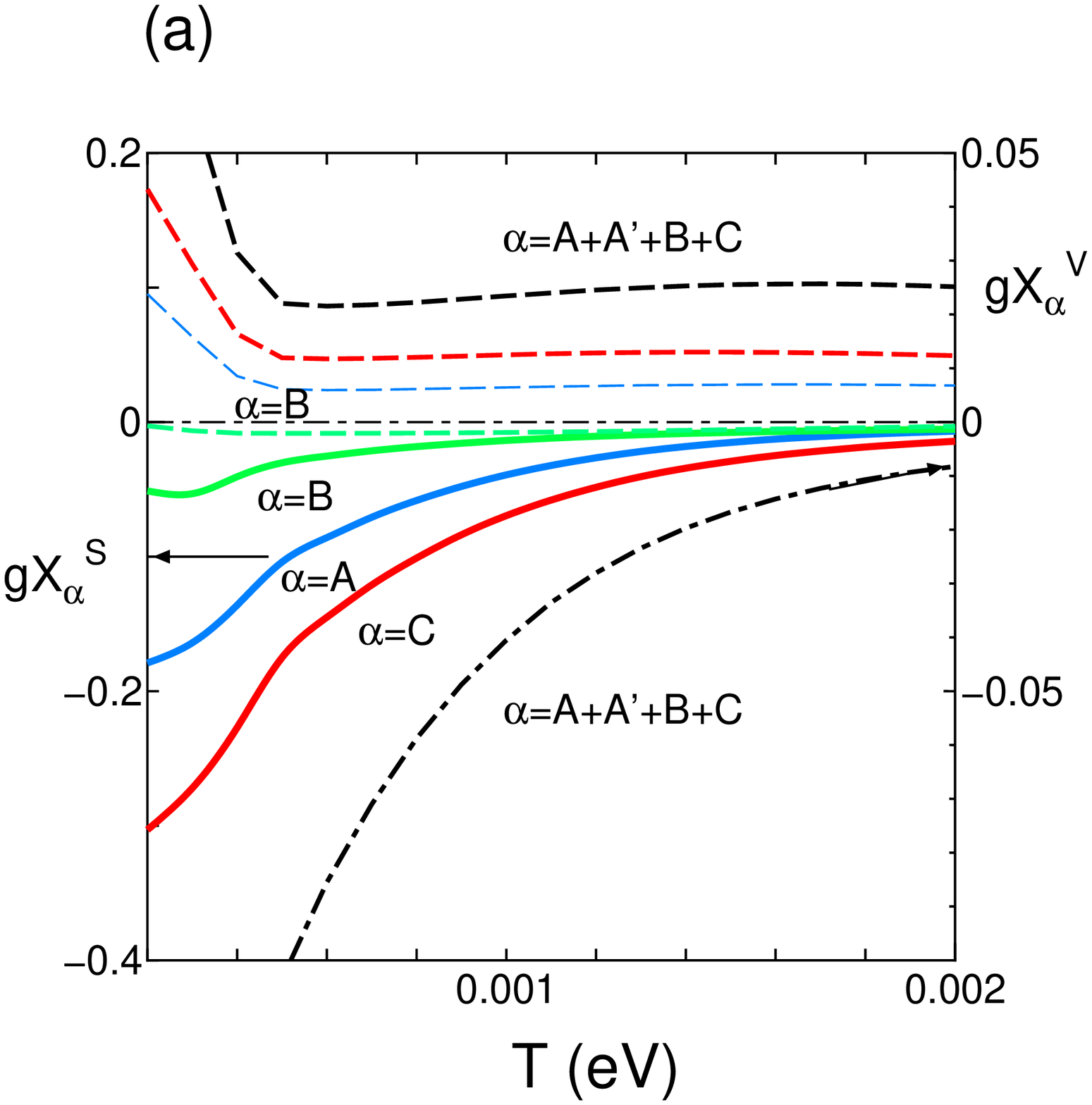}   
\includegraphics[width=7cm]{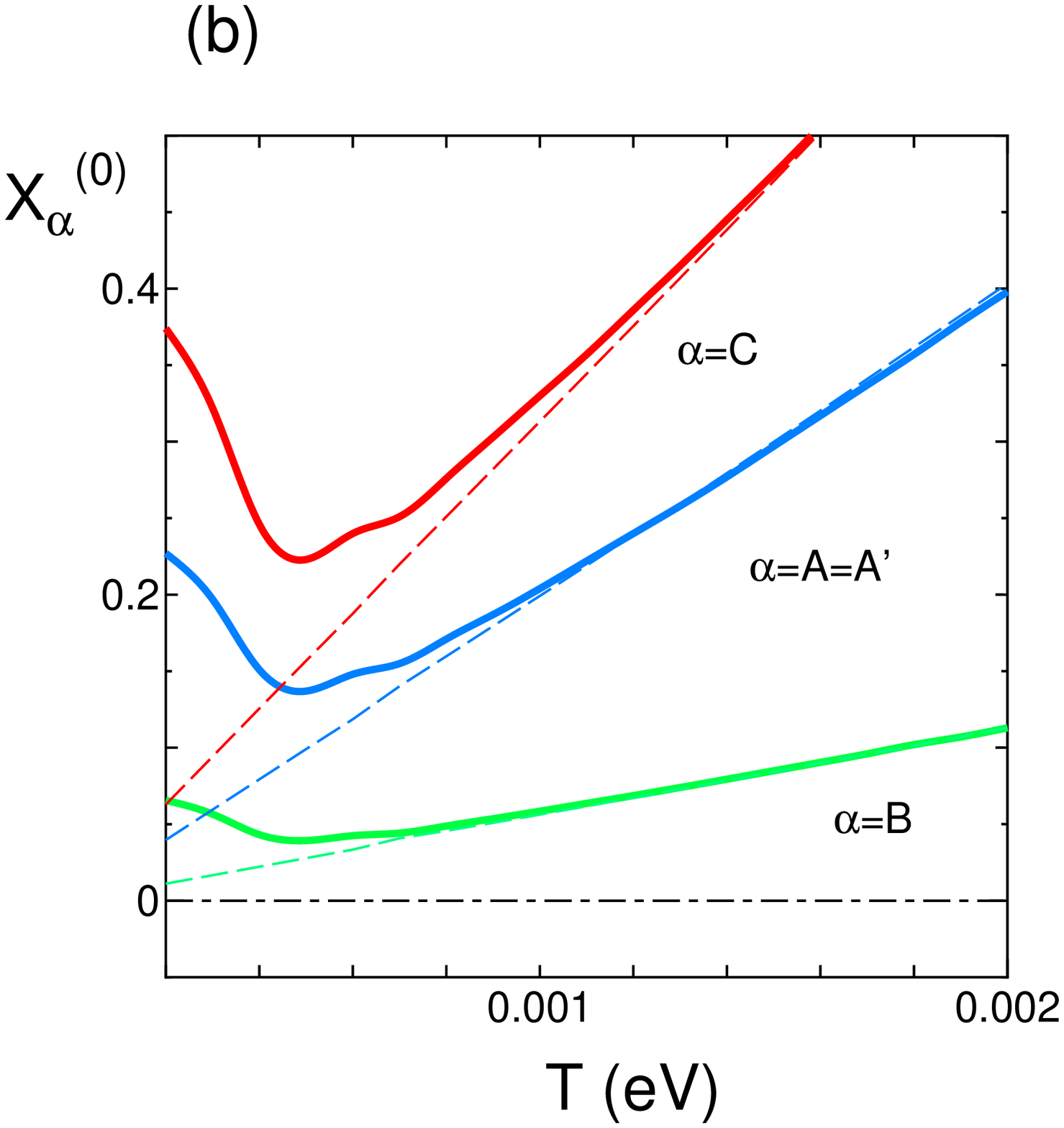}   
  \caption{(Color online) 
(a) $T$ dependence of the self-energy (solid line)  
 and vertex corrections (dashed line) 
  corresponding to Fig.~\ref{fig:fig6}.
 (b) $T$ dependence of the corresponding 
 $\chi_\alpha^{(0)}$ 
  with $\mu \not= 0$ and $g$=0.4.
For $0.0004 < T$, 
  $g \chi_\alpha^{\rm S} (< 0)$ takes a larger magnitude   
  than   that of $g \chi_\alpha^{\rm V} (> 0)$. 
 Thus, the self-energy correction determines  
 the suppression of $\chi_\alpha$  
 in Fig.~\ref{fig:fig6}. 
}
\label{fig:fig7}
\end{figure}

In order to understand the suppression of $\chia$, 
the contributions of self-energy and vertex corrections 
 are examined in Fig.~\ref{fig:fig7}(a).
The effect of the self-energy correction $g\cS$ is much larger than 
that of the vertex correction $g\cV$ at low temperatures of $T < 0.0015$. 
For  $0.0015 < T < 0.002 $, 
  the contribution of $g\cV (>0)$ becomes comparable  with  
 that of $g\cS (<0)$, and then the suppression of $\chia$ becomes small.
At higher temperatures,  it is expected that the vertex correction 
 becomes dominant compared with the self-energy correction, 
 i.e.,  $\chia$ is enhanced compared with  $\chia^{(0)}$.
 Figure \ref{fig:fig7}(b) shows 
 $\chia^{(0)}( =\chia -g\cS_\alpha-g\cV_\alpha)$,  which is always larger than 
 $\chia$ in the absence of the interaction (dashed line) 
 due to $\dmu \not= 0$.
At low temperatures,  $\chia^{(0)}$ is enhanced due to the increase in 
  $|\dmu|$.  
 
\begin{figure}
  \centering
\includegraphics[width=6cm]{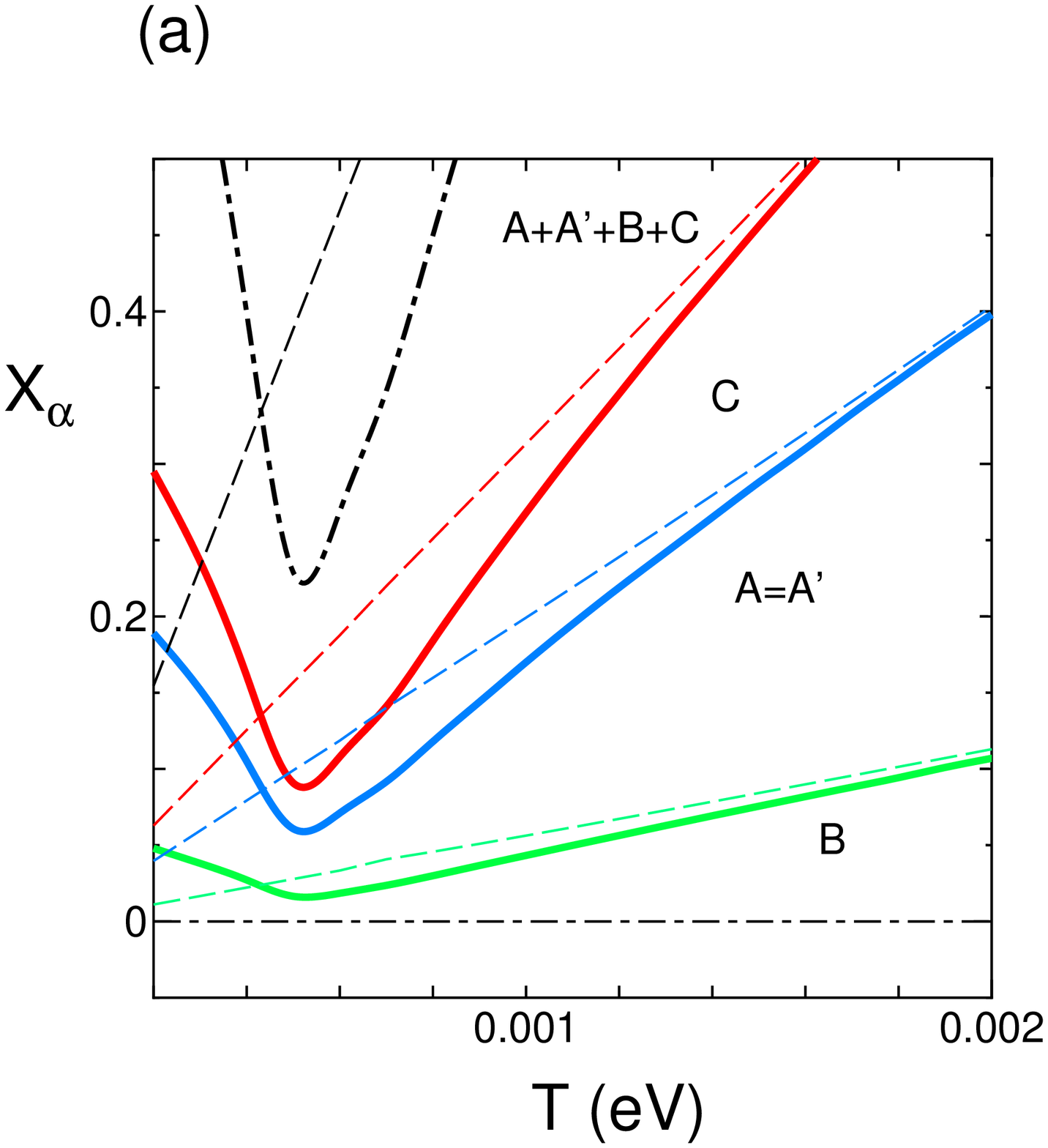}   
\includegraphics[width=6cm]{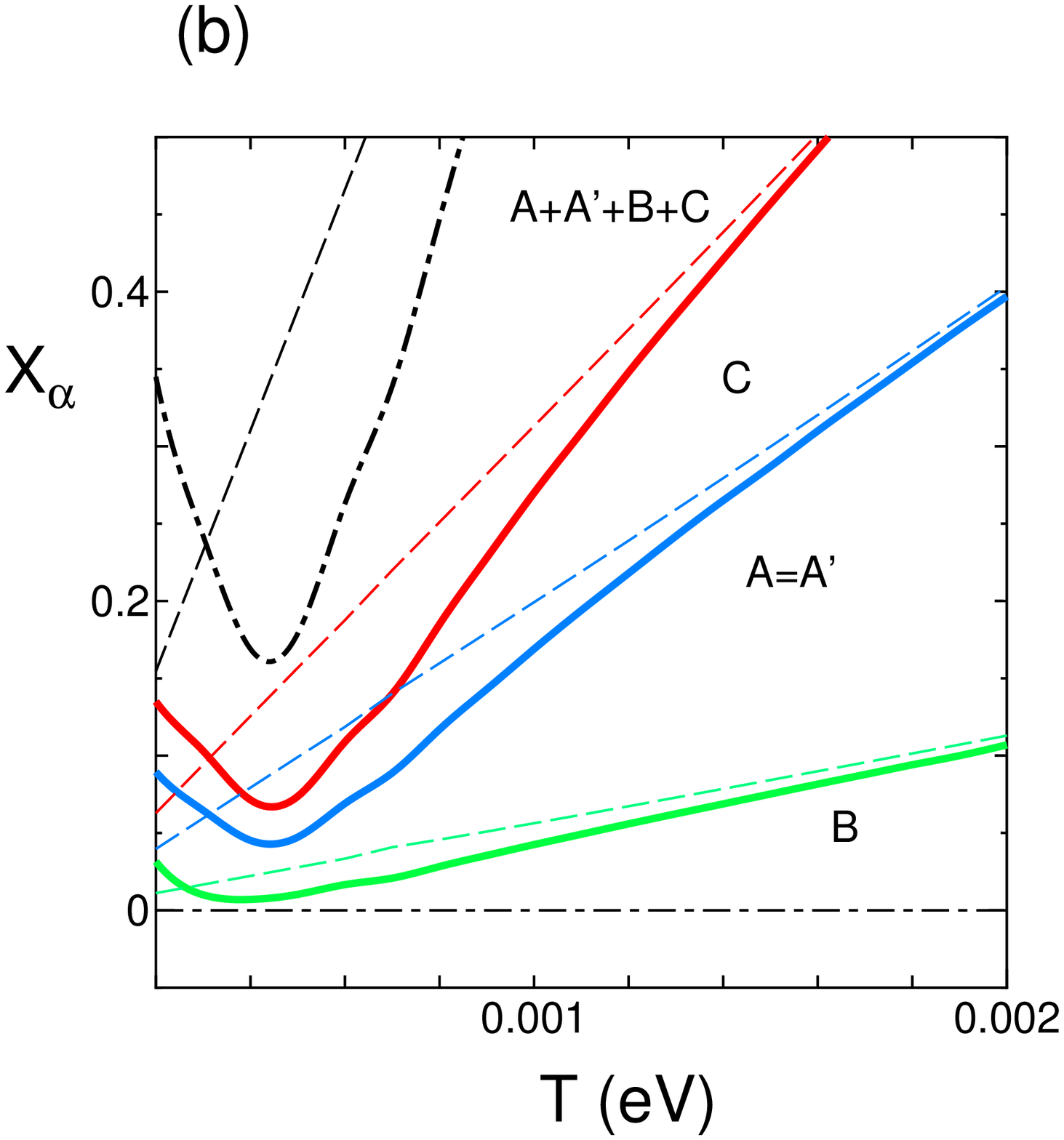} \\  
\hspace{0.5cm}
\includegraphics[width=6cm]{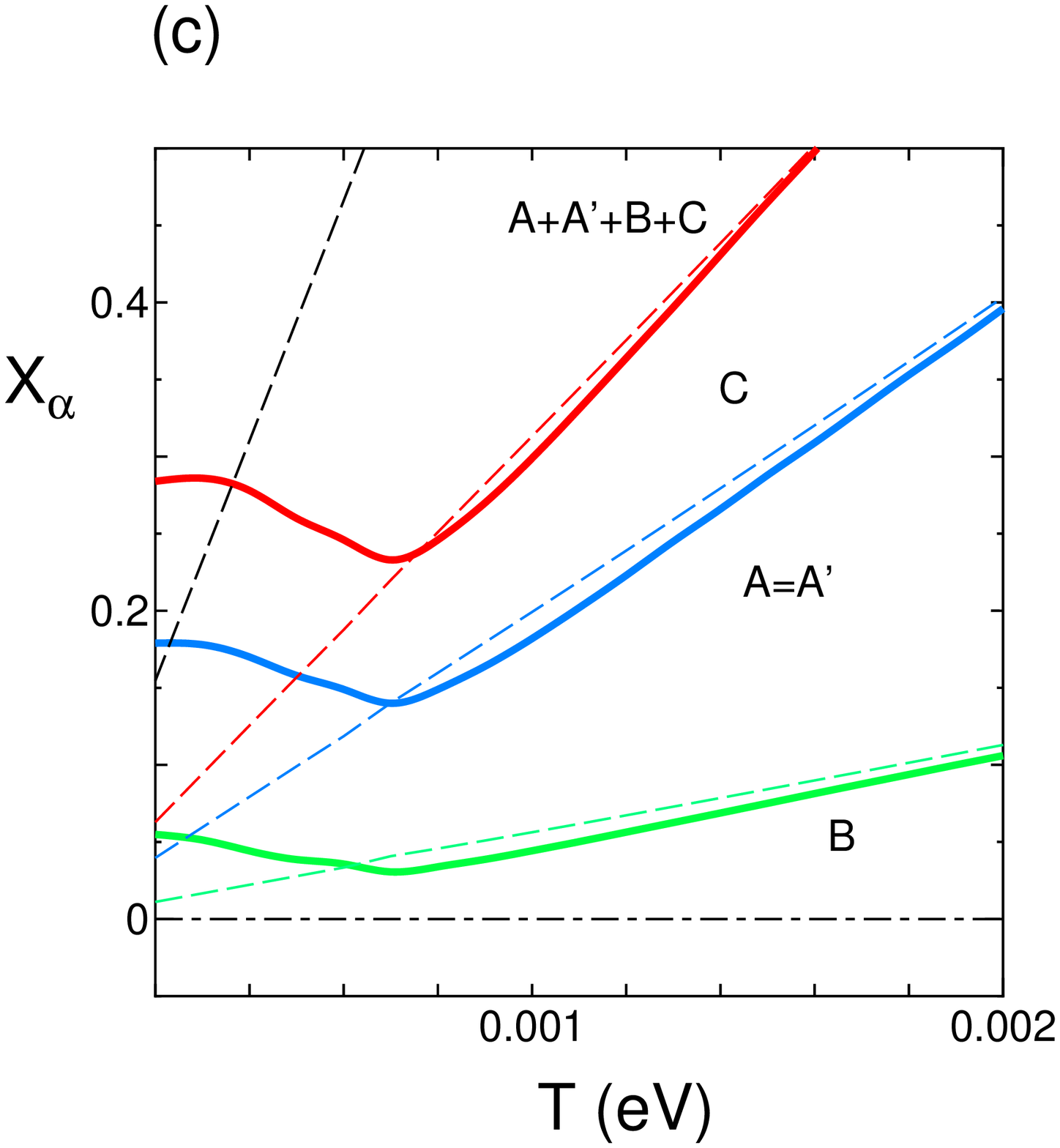}   
\includegraphics[width=6cm]{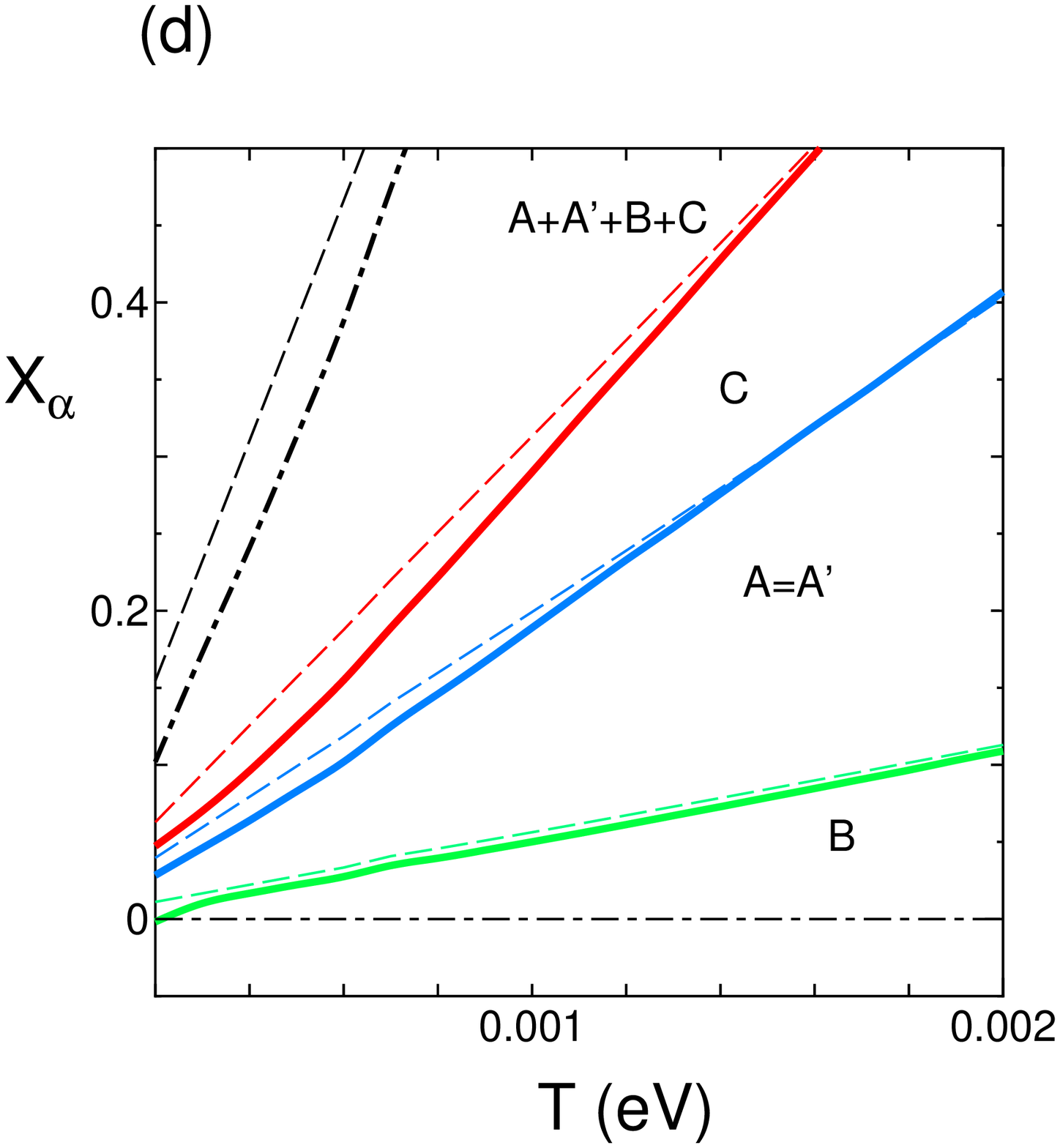}   
  \caption{(Color online)
$T$ dependence of $\chi_{\alpha}$ with $g$=0.04 
 for  $n_d=0.00031$ (a), 
   $n_d=0.00023$ (b),
   $n_d=0$ (c), and 
 $\mu=0$ (d). Notations are the same as in Fig.~\ref{fig:fig6}.
  $\delta \mu$  in (c) is determined self-consistently.
For $0.0008 < T$,  $\delta \chi_\alpha (= \chi_\alpha -\chi_\alpha^{(0)})$
 is negative but small due to the competition of 
 $g \chi_\alpha^{\rm S}$ and $\chi_\alpha^{(0)}$.
 For $n_d$ = 0, 
  $\delta \mu$ becomes  much lower than that of Fig.~\ref{fig:fig4}.
 Thus, the deviation of $\delta \mu$ from zero gives the enhancement 
 of the  DOS.
This enhances the magnitude of both 
 $g \chi_\alpha^{\rm S}$ and $\chi_\alpha^{(0)}$. 
(d) $T$ dependence of $\delta \chi_{\alpha}$, 
  which   is similar  to that in (c), 
 but $\chi_\alpha^{\rm S}$ is much smaller due to the small DOS.  
The  enhancement of $\delta \chi_\alpha$ at low temperatures 
is absent due to $\mu = 0$. 
}
\label{fig:fig8}
\end{figure}

We examine $\chia$ for some other values of $n_d$ 
 for comparison with Fig.~\ref{fig:fig6}.
    Figure 8(a) shows $\chia$ for $n_d= 0.00031$, where 
 the $T$ dependence of $\chia$ is similar but the height  is slightly larger 
 than  that for $n_d=0.00025$. The case of $n_d=0.00031$, 
where   $\dmu (>0)$  for $T < 0.0004$,   
  is  almost on  the boundary of the jump in $\mu$. 
 $\chia$ for $n_d = 0.00028$ 
 is similar but  a jump below the minimum 
 occurs at higher temperatures.
Figure 8(b) shows $\chia$ for $n_d= 0.00023$.  
 The $T$ dependence of $\chia$ is similar to  that of $n_d=0.00025$ 
 but the height is also large. 
 For  $n_d=0.00023$, 
    $\dmu (<0)$  is  slightly  lower than  that of $n_d=0.00025$ owing to 
 being  away from the boundary of the jump of  $\dmu$. 
Thus, there is an optimum value of $n_d$ that gives the lowest $\chia$.
Such $n_d$ is lower and  moderately away from the boundary of the jump. 
The case of $n_d=0$ is shown in Fig.~\ref{fig:fig8}(c)     
   to understand the role of $n_d$  by comparison 
    with  Fig.~\ref{fig:fig6}. 
The height of $\chia$ further increases, 
 but a small suppression ($\chia -\chia^{(0)} < 0)$ for $0.0008 < T$  
 still exists due to competition between 
   the enhancement of  $\chia^{(0)}$ and the decrease in $g \cS$, which 
 occurs for large $|\dmu|$.
However a large  enhancement of $\chia$ is seen at low temperatures 
 since the effect of $|\dmu| \not= 0$ on  $\chia^{(0)}$ is larger 
 than that of $|g \cS|$  at low temperatures. 
Thus, it turns out that  $n_d$ with a moderate magnitude 
has the effect of reducing $\chia$.  
Figure \ref{fig:fig8}(d) shows $\chia$ for $\dmu=0$ and $g=0.04$.
Although the interaction gives $\dmu \not= 0$, 
 the case of $\dmu = 0$ is compared with Fig.~\ref{fig:fig6} 
  to clarify the role of $\dmu$ in $\chia$. 
For $\dmu = 0$, the reduction  given by $\chia - \chia^{(0)} < 0$ 
 still exists but is small. 
A minimum of $\chia$ is absent and $\chia$ 
  decreases monotonically.
 The magnitudes of  $\cS$ and $\cV$ are smaller but 
 their $T$ dependence  is similar to that in Fig.~\ref{fig:fig7}(a) except for low $T (< 0.0005) $.

Thus, the origin of the minimum of $\chia$ is as follows. 
When $|\dmu|$ increases from zero (as found by the presence of $g \not= 0$), 
 the DOS at the chemical potential increases, 
 and the increase in $|\cS|$ becomes much larger than $\cV$, 
 resulting in the large suppression  of $\chia$,
  as seen from Fig.~\ref{fig:fig6}. However, 
  $|\dmu|$ also  
increases $\chia^{(0)}$ at $T < 0.0006$ 
 as shown  in  Fig.~\ref{fig:fig7}.
Such competition gives a minimum of $\chia$ at $T \simeq 0.0004$ 
 in Fig.~\ref{fig:fig6}.

\section{Summary and discussion}

 We examined the NMR shift $\chi_{\alpha}$ 
 at low temperatures of $T < 0.002$ eV 
 for  massless Dirac electrons  
 in the organic conductor $\alpha$-(BEDT-TTF)$_2$I$_3$. 
 The response function was calculated in the presence of 
  the long-range Coulomb interaction,    
where  screenings were taken into account.    
Treating the interaction up to the 
 first order in the perturbation, 
 the  chemical potential in the presence of the doping $n_d$ 
     was  calculated self-consistently,  
 and the response function was calculated  
 for both self-energy and vertex corrections 
    to satisfy the Ward identity. 
The self-consistent solution of $\dmu$ was examined  
 on the plane of $n_d$ and $T$.  
 The suppression of $\chia$ was obtained  using $n_d$  
 close to  the boundary between  $\dmu >0$ and $\dmu <0$  at $T$ = 0. 
We found  a novel fact that 
 both $\chi_{\rm B}^{\rm S} < 0$ and  $\chi_{\rm B}^{\rm V} < 0$.
 The suppression  of $\chia$ originates  from  
 the self-energy correction being dominant over 
  the vertex  correction.
 A minimum of $\chia$  exists at low temperatures. 
 At lower temperatures,  the shift is enhanced  
 due to  $\delta \mu \not= 0$.

Here we compare Fig.~\ref{fig:fig6} with other previous work.
The fact that the sign of the vertex correction $\chi_{\alpha}^{(V)}$ 
 is positive  for $\alpha$ = A and C  but negative for 
 B is compatible with  the model with  
 the  on site-repulsive interaction.\cite{Matsuno2017}  
This suggests a common feature of the vertex correction even though 
 the interaction range is  different between these models. 
 The fact that  $ 0 > g \cABC^{S}$ 
 at low temperatures is consistent with the sign  expected 
 by the calculation of the self-energy of the Green function.
\cite{Kotov2012} 
 The negative sign of $\cABC^{(S)}$ in the present paper  
 is the same as that obtained by calculating the renormalization 
of the velocity of the Dirac cone  in terms of such a Green function.
\cite{Isobe2012} 
In the present calculation,  
 a large suppression of $\chia$ is obtained for a finite  doping ($n_d$)
   with $\dmu \not= 0$,  while 
  suppression is obtained 
  in the absence of doping with $\dmu = 0$ for the case of 
   velocity renormalization.

We note    a reduced model of a 2$\times$2 Hamiltonian
\cite{Kobayashi2007}  
 consisting  of only two bands, the conduction and valence 
 bands,  which are obtained from  
 $\ep_1(\bk)$ and $\ep_2(\bk)$ with $d_{\alpha,1}$ 
 ($\alpha$ = A (= A'), B, C) in Eqs.~(\ref{eq:H_eq}) and (\ref{eq:d_vector}).
Calculating Eqs.~(\ref{eq:n_1}), (\ref{eq:chi_self}), and 
 (\ref{eq:chi_vertex}) with these two bands and  
all the $\alpha$,
 we found that  the difference in the numerical result 
 between the reduced model and the 4$\times$4 Hamiltonian (Eq.~(\ref{eq:H_mat})     is about  3$\%$   
 suggesting the validity of the  effective 2$\times$2 Hamiltonian 
 with  a choice of 
 the base in terms of the Luttinger--Kohn representation.\cite{Kobayashi2007}
The present calculation gives  the NMR shift directly  
 owing to the diagonalization of Eq.~(\ref{eq:H_mat}) 
 for each $\bk$.
Although  the comparison of the intermediate process 
  with   the effective Hamiltonian 
  is complicated due to  the  factors $d_{\alpha,1}$ and $d_{\alpha,2}$
  depending on the choice of the base,
  the same  result of the NMR shift is expected  
  when the components of the  base are  reasonably taken into account.

We took $n_d$ as a parameter 
 to explain the NMR shift.
 The parameter is located slightly away from the first-order transition 
 since, at present, such a transition has not been  found experimentally. 
 The existence of  $n_d (>0)$ is claimed 
 from the Hall conductivity, where a theory
  without interaction\cite{Kobayashi2008}  
 predicted   $n_d \simeq 10^{-6}$  and 
 an experiment\cite{Tajima_PRB} estimated  
$n_d$ = (0.1  -- 1) $\times 10^{-5}$.
The experimental estimation  is reasonable 
 owing to the enhancement of $n_d$ by the interaction.
However, the present choice of $n_d \simeq 10^{-4}$,
  which is  larger than the experimental value,  still remains 
a problem to be resolved in the future.

Finally we discuss the relevance of the present work 
  to the experiment on the  NMR shift in \ETm.
 Site-selective NMR  shows that 
  the electron susceptibility 
 decreases with decreasing $T$ below $< 0.01$ eV   with 
    $\cC > \cA (= \chi_{\rm A'}) > \cB$,\cite{Takahashi2010,Hirata2012}
 where the suppression from the $T$ linear dependence of $\chi_{\alpha}$ 
  is visible  and   the strong suppression of $\cB$ shows 
  a gaplike $T$ dependence.  
 The behavior  at lower temperatures is  as follows.\cite{Hirata2016}
 For $T < 0.005$,  $\chi_B$ becomes  almost zero 
   with a minimum. 
 Also  both $\chi_A$ and $\chi_C$  decrease rapidly.   
  At  $T \simeq 0.002$, all $ \chi_\alpha$ become almost zero.
  This experimental result is compared with our theoretical result 
     of $\chia$ in Fig.~\ref{fig:fig6} 
   ($n_d = 0.00025$ and  $g=0.04$), 
     which shows a large suppression  
      of $\cB$ at low temperatures. 
  Thus,  a  common $T$ dependence is seen  
     for temperatures above the minimum. 
However, the present calculation shows an enhancement 
  at lower temperatures while 
  the experiment shows  monotonic decreases in 
   $\cC$ and $\cA$. 
 Further, the  characteristic temperature in the present calculation 
  is much lower than that in the experiment. 
 Such a difference may be reduced by considering 
  a larger magnitude of $g$.
Another comment is  regarding the chemical potential $\dmu$
 as shown in  Fig.~\ref{fig:fig5}.
 For larger $n_d (=0.0003)$, 
 the $T$ dependence of $\dmu$, which moves from positive to negative, 
    is qualitatively similar to  
  that obtained  theoretically in terms of 
    carrier doping without interaction.
\cite{Kobayashi2008}
In fact, 
    such a  change of the sign, which gives rise to the change in   
  the Hall coefficient, 
    was verified 
     by an experiment on the Hall conductivity.\cite{Tajima_PRB}

\acknowledgements
The author thanks H. Fukuyama  for the suggestion of the problem and 
 valuable comments, and  A. Kobayashi for useful discussions.
This work was supported by JSPS KAKENHI Grant Numbers JP15H02108 and JP26400355.
\newpage

\appendix

\section{Effective interaction}
We analytically calculate the screening constant for the bare Coulomb interaction
 (Eq.~(\ref{eq:H_int})) using an effective 2$\times$2  Hamiltonian 
\cite{Kobayashi2007,Kobayashi2008} 
for the Dirac cone around the Dirac point $\bkD$,   given by 
\begin{eqnarray}
H_{\rm eff}&=& 
\begin{pmatrix}
v \tilde{k}_y + \lambda v \tilde{k}_x & v \tilde{k}_x  \\
v \tilde{k}_x  & - v \tilde{k}_y + \lambda v \tilde{k}_x
\end{pmatrix} \; ,
\label{H:Koba_Hall}
\end{eqnarray}
 where $\tilde{\bk} = (\tilde{k}_x ,\tilde{k}_y) =  \bk -\bkD$  
     with the Dirac point $\bkD$. 
For simplicity we rewrite as $\tilde{\bk} \rightarrow \bk$. 
 The eigenvalue of Eq.~(\ref{H:Koba_Hall})
 is given by  $\xi_{\gamma, \bk}= v \lambda k_x +\gamma v |\bk|$ 
  with $(\gamma = \pm)$.
Equation (\ref{H:Koba_Hall}) describes the Dirac cone with tilting parameter 
 $\lambda$, where 
 the  $k_x$ axis is taken as the tilting direction.
The poralization  function 
of  Eq.~(\ref{H:Koba_Hall}), which is 
given by the  density-density response function, 
 is written as\cite{Nishine2010}  
\begin{eqnarray}
 \Pi(\bm{q},\dmu,T) & = & 
 - \frac{2}{N^2} \sum_{\gamma,\gamma'}\sum_{\bk} \frac{1 
 + \gamma \gamma' (\bk \cdot \bk')/|\bk||\bk'|}{2} \times 
\frac{f(\xi_{\gamma,\bk})-f(\xi_{\gamma',\bk'})}
{-\xi_{\gamma',\bk'} + \xi_{\gamma, \bk}}
  \; ,
 \label{eq:Pi_qmT}
 \end{eqnarray} 
where $\bk' = \bk+\bm{q}$ and $f(\xi)=1/(\exp[(\xi-\dmu)/T]+1)$.
Using Eq.~(\ref{eq:Pi_qmT}), 
the effective Coulomb interaction within the RPA is written as  
\begin{eqnarray}
v_{\bq,{\rm  eff}} &=& 
    \frac{v_q}{1+\qTF/q} 
       = \frac{g l}{|\bq|+\qTF}   \; ,
 \label{eq:ap_veff}
\end{eqnarray}
where 
\begin{eqnarray}
& & v_q   = \frac{v_q^0/\ep_2}{1+v_q^0\Pi(\bm{q},0,0)} =  \frac{v_q^0}{\ep}
 \equiv  \frac{g l}{q}
\; ,
\label{eq:ap_vq} 
     \\
 & & \Pi(\bq,0,0) = \frac{q}{2\pi v}
 \left< \frac{1}{\sqrt{1 - \lambda^2 \cos^2{\theta_q}}}
            \right>_{\theta_q} \; ,
 \label{eq:pi_0} 
 \end{eqnarray}
 $g =2 \pi e^2/(\ep l)$, 
 $v_q^0 = 2 \pi e^2 / q$, $q=|\bq|$, and $l$ is the lattice constant.
 $\ep = \ep_1 \ep_2$.  $\ep_1 (= 1+v_q^0\Pi(\bm{q},0,0))$\cite{Nishine2010} 
 is the intralayer dielectric constant and   
 $\ep_2$ denotes the interlayer dielectric constant,  taken 
 as $\simeq$ 5. 
 $<>_{\theta}$ denotes the average over the  angle $\theta=\theta_q$, 
 which denotes  the angle between $\bm{q}$ and  the tilted axis of the Dirac cone with  tilting parameter $\lambda$.
Equation (\ref{eq:pi_0} ) is multiplied by 4 due to 
 the freedom of the spin and 
 valley.  
 In Eq.~(\ref{eq:ap_veff}), the denominator, $1+\qTF/q$, 
  is an interpolation formula used 
to describe the crossover  between small $q (<< \qTF)$ and large $q (>> \qTF)$. 
This gives a reasonable result compared with the exact one.\cite{Nishine2010}
 Assuming  only the intralyer screening
 due to $\dmu (\not= 0)$,  $\qTF$ is written as 
 \begin{eqnarray}
 \qTF &= & q v_q \Pi(0,\dmu,T)
  \simeq q v_q \times 
    \frac{4 (| \dmu | + T)}{2 \pi v^2 (1 - \lambda^2)^{3/2}}
 \; ,
 \label{eq:pi_TF}
 \end{eqnarray}
 which is the Thomas--Fermi screening including temperature.  
Thus, $\qTF$ is  estimated as  
\begin{eqnarray}
\qTF & \simeq & \frac{4 e^2 / v}{\ep (1 - \lambda^2)^{3/2}} \times \frac{|\dmu| + T}{v}
 \simeq  2.5 \times  \frac{|\dmu| + T}{v} \; ,
\label{eq:qTF}
 \end{eqnarray} 
 where $g =  0.21 /\ep_2$, 
$\ep_1 = 1+ v_q^0 \Pi (\bq, 0,0) \simeq 40 $.
In deriving Eq.~(\ref{eq:qTF}), we used 
 the parameters  $\lambda \simeq 0.8$, 
$e^2/v =27.2$,  
$2 \pi e^2/l = 8.5$ eV, 
$v/l = 0.05$ eV,
$<(1-\lambda^2 \cos^2 \theta)^{-1/2}>_\theta =1.43$, and 
$(1-\lambda^2)^{-3/2}=4.62$. 
Note that $g=0.04$ corresponds to $\ep_2 \simeq 5$. 


\section{Number density}
Using the matrix, 
$S(1/T)$=$T_{\tau} \exp [- \int_{0}^{1/T} H_{\rm int}(\tau) {\rm d} \tau)] $,
\cite{Abrikosov} 
 where 
 $T_{\tau}$ is the ordering operator of the imaginary time ($\tau$) 
 and $H_{\rm int}(\tau) = 
{\rm e}^{(H_0-\mu) \tau} H_{\rm int} {\rm e}^{-(H_0-\mu) \tau}$,  
we calculate 
the density and response functions up to the first order in $H_{\rm int}$. 

The number density per unit cell and  per spin is calculated from 
\begin{eqnarray}
  \lim_{\tau \rightarrow - 0} 
 \frac{1}{N} \sum_{i}  \sum_{\alpha} 
\frac{ - \langle T_{\tau}( \psi_{i,\alpha}(\tau) \psi_{i,\alpha}(0)^{\dagger} S(1/T)) \rangle_0}{ \langle S(1/T) \rangle_0 } 
 \simeq  n^{(0)} + g n^{(1)} 
\; ,
 \label{eq:a1}
 \end{eqnarray} 
where 
 $\langle \; \rangle_0$ denotes the thermal average on $H_0$.
From Eq.~(\ref{eq:H_eq}) with  $\psi_{\bk \alpha} 
 = \sum_{\gamma} d_{\alpha \gamma}(\bk) 
     \psi_{\bk \gamma}$, 
 the density of the zeroth order shown in Fig.~\ref{fig:density}(a)
 is calculated as 
\begin{eqnarray}
n^{(0)} &=& \frac{1}{N} \sum_{i}  \sum_{\alpha} 
   \langle \psi_{i,\alpha}^{\dagger} \psi_{i,\alpha} \rangle_0
 =  \frac{1}{N}  \sum_{\bk}  \sum_{\alpha} 
\langle \psi_{\bk \alpha}^{\dagger} 
 \psi_{\bk \alpha} \rangle_0
    \nonumber \\
&& =  \frac{1}{N} \sum_{\gamma}  \sum_{\bk} \sum_{\alpha}
    d_{\alpha \gamma}(\bk)^*d_{\alpha \gamma}(\bk) \times 
 \langle \psi_{\bk \gamma}^{\dagger} \psi_{\bk \gamma} \rangle_0 
  \nonumber \\ 
& & = 
\frac{1}{N}  \sum_{\gamma} \sum_{\bk}
T \sum_{n} 
 G(n,\ep_\gamma(\bm{k}))
  = 
\frac{1}{N} \sum_\gamma  \sum_{\bk}
 f(\ep_\gamma(\bm{k}))  \; .
 \label{eq:a2}
 \end{eqnarray} 
The Green function is given  by 
  $G(n,\ep_\gamma(\bm{k}))$= 
 $\int ( -T_{\tau} \langle \psi_{\bk \gamma}(\tau) 
 \psi_{\bk \gamma}(0)^{\dagger} \rangle 
{\rm e}^{- i \omega_n \tau} {\rm d}\tau$
 =  ($i\omega_n + \mu  -\ep_\gamma(\bm{k}))^{-1}$, where  
$\omega_n (= (2n+1)\pi T)$ is the Matsubara frequency 
   with $n$ being an integer 
 and 
  $T \sum_n G(n, \ep_{\gamma}(\bk)) = f(\ep_{\gamma}(\bk)) 
= 1/( \exp [(\ep_{\gamma}(\bk) - \mu )/T]+1)$.

The density of the first order is calculated as (Fig.~\ref{fig:density}(b))
\begin{eqnarray}
g n^{(1)} &=& -g  \frac{T^2}{N^2l^2} \sum_{n,n'}  \sum_{\bk,\bq}
 \sum_{\gamma_1, \gamma_2,\gamma_3}  \sum_{\alpha',\beta'}
 \frac{1}{|\bq|+\qTF}
 G(n,\ep_{\gamma_1}(\bk))  G(n,\ep_{\gamma_2}(\bk))  G(n',\ep_{\gamma_3}(\bk-\bq))
    \nonumber \\
& &\times 
  d_{\alpha \gamma_1}^{*}(\bm{k})d_{\alpha' \gamma_1}(\bm{k}) 
   d_{\alpha'\gamma_3}^{*}(\bm{k}-\bm{q})
    d_{\beta'\gamma_3}(\bm{k}-\bm{q})
  d_{\beta'\gamma_2}^{*}(\bm{k})d_{\alpha \gamma_2}(\bm{k}) 
 \nonumber \\ 
  &=& -g  \frac{T^2}{N^2l^2} \sum_{n,n'}  \sum_{\bk,\bq} 
 \sum_{\gamma_1, \gamma_3}  \sum_{\alpha',\beta'} \frac{1}{|\bq|+\qTF}
 G(n,\ep_{\gamma_1}(\bk))^2 G(n',\ep_{\gamma_3}(\bk-\bq))
    \nonumber \\
& &\times d_{\alpha'\gamma_1}(\bm{k})d_{\beta'\gamma_1}^{*}(\bm{k}) 
   d_{\alpha'\gamma_3}^{*}(\bm{k}-\bm{q})
    d_{\beta'\gamma_3}(\bm{k}-\bm{q})
 \nonumber \\ 
 & = &  - g  \frac{1}{N^2l^2}  \sum_{\bk,\bq} 
 \sum_{\gamma_1, \gamma_3}   \sum_{\alpha',\beta'}
\frac{1}{|\bq|+\qTF} \times 
       \frac{\partial f(\ep_{\gamma_1}(\bm{k}))}
{\partial \ep_{\gamma_1}(\bm{k})} 
\times 
f(\ep_{\gamma_3}(\bm{k}-\bm{q}))
  \nonumber \\
& & \times d_{\alpha'\gamma_1}(\bm{k})d_{\beta'\gamma_1}^{*}(\bm{k}) 
   d_{\alpha'\gamma_3}^{*}(\bm{k}-\bm{q})
    d_{\beta'\gamma_3}(\bm{k}-\bm{q}) 
\; .
 \label{eq:a3}
 \end{eqnarray} 
Equation (\ref{eq:a3})  leads to Eq.~(\ref{eq:n_1}).
Note that  $g n^{(1)} > 0$ 
 since  $ - \partial f(\ep) / \partial \ep > 0$   and $f(\ep) > 0$. 

At $T$=0, Eq.~(\ref{eq:n_1}) is examined 
  using an effective 2$\times$2 Hamiltonian (Appendix A)
 with $\gamma = \pm$ and  tilting parameter $\lambda$, 
   where  the Dirac cone is tilted with  maximum velocity 
 $v(1+\lambda)$ and minimum velocity  $v(1-\lambda)$. 
Equation  (\ref{eq:n_1}) is  calculated as 
\begin{eqnarray}
   n^{(1)} &= &  \frac{|\dmu|l^3}{2\pi^2}
       \int_{0}^{k_c} d y \; y \sum_{\gamma_3=\pm}
  \left< \frac{  f(\ep_{\gamma_3}(\bk'))
  <\gamma_3(\bk') | \gamma_1(\bk)> |^2/v(\theta)^2
}{\sqrt{k_{\mu}^2 + y^2 -  2 y k_{\mu}  \cos (\theta - \theta')}
 +q_{\rm TF}(\dmu,0)
 }\right>_{\theta,\theta'} \; ,
 \label{eq:n_T=finite}
 \end{eqnarray}
 where 
   $\delta \mu = \mu -  \mu_0$ with $\mu_0$ given by  
  $\ep(\bm{k}_D)$ at $T=0$. 
$\gamma_1$ = $+$ or $-$, $<>_\theta$ denotes an average
  with respect to $\theta$, 
$y=k'$, 
 $k_c (>> \dmu/v)$ is the momentum cutoff of the Dirac cone  
$v_{\theta} = v (1+\lambda \cos \theta)$, 
 $\bk=k(\cos \theta, \sin \theta)$, 
 $\bk'=k'(\cos \theta', \sin \theta')$, 
and $k_{\mu}= |\dmu| / v(\theta)$.
 For  $\lambda$ = 0.8  and $v/l \simeq$ 0.05,   
 the numerical estimation gives  
$n^{(1)} = C_1 |\dmu|$ with $C_1 \simeq 12$ (eV)$^{-2}$.

\section{Response function}
The NMR shift at the $\alpha$ site is obtained from 
\begin{eqnarray}
\chi_{\alpha} = \sum_{\beta} \chi_{\alpha \beta}
  \simeq \chi_\alpha^{(0)} + g \chi_{\alpha}^{\rm S} 
 + g \chi_{\alpha}^{\rm V} \; ,
 \label{eq:a4}
 \end{eqnarray} 
 where $\chi_{\alpha \beta}$ is the response function 
    between the $\alpha$  and $\beta$ sites, which is calculated by
\cite{Abrikosov}  
\begin{eqnarray}
\chi_{\alpha \beta} & = & \frac{1}{N} \sum_{\bk}
  \int_{0}^{1/T}
\frac{  \left< T_{\tau} ( 
  \psi_{\bk \alpha}^{\dagger}(\tau)  \psi_{\bk \alpha}(\tau)
    \psi_{\bk \beta}^{\dagger}(0) \psi_{\bk \beta}(0) S(1/T))\right>_0
 }{
   \langle S(1/T) \rangle_0}
 \; {\rm e}^{i \omega_n \tau} {\rm d} \tau 
\vert_{i\omega_n \rightarrow + i0}
  \; . 
 \label{eq:chi_ab}
 \end{eqnarray} 
We took  $2 \mu_{\rm B}^2$  as unity  with 
 $\mu_{\rm B}$ being the Bohr magneton. 
Equation (\ref{eq:chi_ab}) is calculated by expanding $S(1/T)$,  
 in terms of $H_{\rm int}$ where 
 the zeroth order gives $\chi_\alpha^{(0)}$  and 
 the first order gives $g \chi_{\alpha}^{\rm S} 
 + g \chi_{\alpha}^{\rm V}$.
In the second-order terms, there is the  A--L contribution 
 whose  diagram reduces   to a disconnected diagram\cite{Abrikosov}
  in the absence of $H_{\rm int}$.
 Such a contribution, which is  added to Eq.~(\ref{eq:a4}) to satisfy 
 the Ward identity\cite{Ward}  
  for the RPA given by Eq.~(\ref{eq:veff_m}), 
 vanishes in the present case due to 
  the summation of $\hat{m}_{j\beta}$ in Eq.~(\ref{eq:response}) with respect to $\beta$.

From Fig.~\ref{fig:chi}(a), the zeroth order is calculated as 
\begin{eqnarray}
 \chi_{\alpha}^{(0)} &= &  
 - \frac{T}{N} \sum_n \sum_{\bk} \sum_{\gamma, \gamma'} \sum_{\beta}
 G(n,\ep_{\gamma}(\bk) )  G(n,\ep_{\gamma'}(\bk)) 
 d_{\alpha \gamma}^{*}(\bk) d_{\beta \gamma}(\bk)
 d_{\beta \gamma'}^{*}(\bk) d_{\alpha \gamma'}(\bk) 
\nonumber \\
 & & =
  - \frac{1}{N}\sum_{\bm{k}} \sum_{\gamma, \gamma'} \sum_{\beta}  
  \frac{f(\ep_{\gamma}(\bk)) - f(\ep_{\gamma'}(\bk))} 
  {\ep_{\gamma}(\bk)-\ep_{\gamma'}(\bk)}
 d_{\alpha \gamma}^{*}(\bk) d_{\beta \gamma}(\bk)
 d_{\beta \gamma'}^{*}(\bk) d_{\alpha \gamma'}(\bk) 
  \nonumber \\ 
& &= 
- \frac{1}{N} \sum_{\bm{k},\gamma} \frac{\partial f(\ep_{\gamma}(\bm{k}))}
 {\partial \ep_{\gamma}(\bm{k})}
   d_{\alpha \gamma}^{*}(\bm{k})d_{\alpha \gamma}(\bm{k})
 \; .
 \label{eq:a6}
 \end{eqnarray} 
  
  In Eq.~(\ref{eq:a6}), we used the identity 
\begin{eqnarray}
  \sum_{\beta}  d_{\beta \gamma}(\bk) d_{\beta \gamma'}^{*}(\bk)
 = \delta_{\gamma, \gamma'} 
  \; ,
\label{eq:identity_1}
 \end{eqnarray} 
which is also applied in the following calculation 
 of   $ g \chi_{\alpha}^{S}$  and $ g \chi_{\alpha}^{V}$.

The first order consists of 
 the self-energy correction $g \chi_{\alpha}^{S}$ 
 and the vertex correction $g \chi_{\alpha}^{V}$. 
From Figs.~\ref{fig:chi}(b) and \ref{fig:chi}(c),
the self-energy correction is calculated as 
\begin{eqnarray}
g \chi_{\alpha }^{S}&=& 
    \frac{gT^2}{N^2l^2} \sum_{n,n'} \sum_{\bk, \bq} \frac{1}{|\bq|+\qTF}
 \sum_{\gamma_1,\gamma_2,\gamma3,\gamma4}
\sum_{\alpha',\beta'} \sum_{\beta}
  G(n,\ep_{\gamma_1}(\bk)) G(n',\ep_{\gamma_4}(\bk-\bq))
   G(n,\ep_{\gamma_3}(\bk)) G(n,\ep_{\gamma_2}(\bk))
 \nonumber \\
& & 
 \times 
   d_{\alpha \gamma_1}^{*}(\bm{k})   d_{\alpha' \gamma_1}(\bm{k})
   d_{\alpha' \gamma_4}^{*}(\bm{k}-\bm{q}) d_{\beta' \gamma_4}(\bm{k}-\bm{q})
   d_{\beta' \gamma_3}^{*}(\bm{k})  d_{\beta \gamma_3}(\bm{k})
   d_{\beta \gamma_2}^{*}(\bm{k})  d_{\alpha \gamma_2}(\bm{k}) 
 +  (1 \leftrightarrow   2  ) \; .
  \label{eq:self_ene_1}
   \\
    \nonumber 
\end{eqnarray} 
Using Eq.~(\ref{eq:identity_1}) and 
 the partial fraction decomposition in terms of  
  $G(n,\ep_{\gamma})$, 
\begin{eqnarray}
g \chi_{\alpha }^{S} & = &  \frac{g}{N^2l^2}   \lim_{3 \to 2} \sum_{\bk, \bq} \frac{1}{|\bq|+\qTF}
 \sum_{\gamma_1,\gamma_2,\gamma3,\gamma4} \sum_{\alpha',\beta'}
  \nonumber \\ && 
  \left( \frac{f_1}{(\ep_1-\ep_2)(\ep_1-\ep_3)}
       + \frac{f_2}{(\ep_2-\ep_3)(\ep_2-\ep_1)} 
       + \frac{f_3}{(\ep_3-\ep_1)(\ep_3-\ep_2)} 
\right)  
 \nonumber \\
& & 
 \times 
   d_{\alpha \gamma_1}^{*}(\bm{k})   d_{\alpha' \gamma_1}(\bm{k})
   d_{\alpha' \gamma_4}^{*}(\bm{k}-\bm{q}) d_{\beta' \gamma_4}(\bm{k}-\bm{q})
   d_{\beta' \gamma_2}^{*}(\bm{k})  d_{\alpha \gamma_2}(\bm{k}) 
         \times \left( f(\ep_4)/2 \right) 
     +  (1 \leftrightarrow   2  )
  \nonumber \\
 & =&  \frac{g}{2N^2l^2}
   \sum_{\bm{k},\bm{q}} \sum_{\gamma_1, \gamma_2,\gamma_4} 
\sum_{\alpha',\beta'}  \frac{1}{|\bq|+\qTF} \times
\left(
\frac{f_1-f_2}{(\ep_1-\ep_2)^2} 
+ \frac{1}{\ep_2-\ep_1} \frac{\partial f_2}{\partial \ep_2}
\right)
 \times \left( f(\ep_4)  \right) 
  +  (1 \leftrightarrow   2  ) 
         \nonumber \\
 & &  \times 
   d_{\alpha \gamma_1}^{*}(\bm{k}) d_{\alpha \gamma_2}(\bm{k})
   d_{\alpha' \gamma_1}(\bm{k}) d_{\beta' \gamma_2}^{*}(\bm{k})
   d_{\alpha' \gamma_4}^{*}(\bm{k}-\bm{q}) d_{\beta' \gamma_4}(\bm{k}-\bm{q})
         \; ,
 \label{eq:self_ene_2}
 \end{eqnarray} 
 which leads to Eq.~(\ref{eq:chi_self}).
 $f_1=f(\ep_1)$, $f_2=f(\ep_2)$, $f_4=f(\ep_4)$, 
 $\ep_1= \ep_{\gamma_1}(\bm{k})$, $\ep_2=\ep_{\gamma_2}(\bm{k})$, and 
 $\ep_4= \ep_{\gamma_4}(\bm{k}-\bm{q})$. 
Since $\partial^2 f(\ep)/ \partial \ep^2 >0$ and $f(\ep) > 0$, 
one finds that $\sum_{\alpha} g \chi_{\alpha }^{S} < 0$.

Applying a method similar to Eq.~(\ref{eq:self_ene_1}), 
the vertex correction shown by  Fig.~\ref{fig:chi}(d)
is calculated as 
\begin{eqnarray}
 g \chi_{\alpha }^{V}&=&   \frac{gT^2}{N^2l^2} \sum_{n,n'} \sum_{\bk, \bq} 
 \frac{1}{|\bq|+\qTF}
 \sum_{\gamma_1,\gamma_2,\gamma3,\gamma4}
\sum_{\alpha',\beta'} \sum_{\beta}  \nonumber \\
 & &
  G(n,\ep_{\gamma_1}(\bk)) G(n',\ep_{\gamma_3}(\bk-\bq))
   G(n',\ep_{\gamma_4}(\bk-\bq)) G(n,\ep_{\gamma_2}(\bk))
 \nonumber \\
& & 
 \times 
   d_{\alpha \gamma_1}^{*}(\bm{k})   d_{\alpha' \gamma_1}(\bm{k})
   d_{\alpha' \gamma_3}^{*}(\bm{k}-\bm{q}) d_{\beta \gamma_3}(\bm{k}-\bm{q})
   d_{\beta \gamma_4}^{*}(\bm{k}-\bq)  d_{\beta' \gamma_4}(\bm{k}-\bq)
   d_{\beta' \gamma_2}^{*}(\bm{k})  d_{\alpha \gamma_2}(\bm{k}) 
  \nonumber \\
 &=& 
    \frac{g }{N^2l^2}   \lim_{4 \to 3} \sum_{\bk, \bq} \frac{1}{|\bq|+\qTF}
 \sum_{\gamma_1,\gamma_2,\gamma3,\gamma4}  \sum_{\alpha',\beta'}
  \frac{f_1-f_2}{\ep_1-\ep_2} \times \frac{f_3-f_4}{\ep_3-\ep_4}
  \nonumber \\
& & 
 \times 
   d_{\alpha \gamma_1}^{*}(\bm{k})   d_{\alpha' \gamma_1}(\bm{k})
   d_{\alpha' \gamma_3}^{*}(\bm{k}-\bm{q})
 d_{\beta' \gamma_4}(\bm{k}-\bq)
   d_{\beta' \gamma_2}^{*}(\bm{k})  d_{\alpha \gamma_2}(\bm{k}) 
       \; ,
 \label{eq:a8}
 \end{eqnarray} 
 which leads to Eq.~(\ref{eq:chi_vertex}).
 $f_1=f(\ep_1)$, $f_2=f(\ep_2)$, $f_3=f(\ep_3)$, 
 $\ep_1= \ep_{\gamma_1}(\bm{k})$, $\ep_2=\ep_{\gamma_2}(\bm{k})$, and 
 $\ep_3= \ep_{\gamma_3}(\bm{k}-\bm{q})$. 
 In the last equality, we used the fact that 
  the summation with respect to $\beta$ gives $\ep_3 = \ep_4$. 
Note that 
$[(f_1-f_2)/(\ep_1-\ep_2)] \times 
[(f_3-f_4)/(\ep_3-\ep_4)] > 0 $ 
 due to  $f(\ep)$ being a monotonically decreasing function with $\ep$ 
 and  that $ \sum_{\alpha}d_{\alpha \gamma_1}^{*}(\bm{k}) \cdots   d_{\alpha \gamma_2}(\bm{k}) > 0$, suggesting 
  $\sum_\alpha g \chi_{\alpha }^{V} > 0$.



\end{document}